\documentclass{article}
\usepackage{arxiv}

\renewcommand{\headeright}{Accepted Manuscript}
\renewcommand{\undertitle}{Accepted Manuscript}

\usepackage[utf8]{inputenc}
\usepackage[T1]{fontenc}
\usepackage{hyperref}
\usepackage{url}
\usepackage{natbib}

\usepackage{tabularx}
\usepackage{booktabs}
\usepackage{array}
\usepackage{multirow}
\usepackage{graphicx}
\usepackage{pifont}
\usepackage{soul}

\usepackage{xcolor,stfloats}
\usepackage{makecell} 
\sethlcolor{yellow}

\renewcommand{\arraystretch}{1.4}

\newcommand{\revised}[1]{\textcolor{black}{#1}}

\title{WeldAR: Augmenting Live Hands-On Training with In-Situ Guidance for Novice Learners}

\renewcommand{\shorttitle}{WeldAR: Augmenting Live Hands-On Training}

\author{
  \href{https://orcid.org/0009-0008-2242-461X}{\includegraphics[scale=0.06]{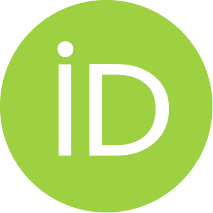}\hspace{1mm}Chuhan (Franklin) Xu}
  \thanks{Equal contribution.} \\
  Carnegie Mellon University\\
  Pittsburgh, Pennsylvania, USA \\
  \texttt{flinkxuchuhan@gmail.com} \\
  \And
  \href{https://orcid.org/0009-0001-6301-8529}{\includegraphics[scale=0.06]{orcid.pdf}\hspace{1mm}Lia Sparingga Purnamasari} \\
  School of Design, Carnegie Mellon University\\
  Pittsburgh, Pennsylvania, USA \\
  \texttt{lpurnama@andrew.cmu.edu} \\
  \And
  \href{https://orcid.org/0000-0002-8470-2709}{\includegraphics[scale=0.06]{orcid.pdf}\hspace{1mm}Zhenfang Chen}\footnotemark[1] \\
  Carnegie Mellon University\\
  Pittsburgh, Pennsylvania, USA \\
  \texttt{zhenfanc@andrew.cmu.edu} \\
  \And
  \href{https://orcid.org/0000-0001-7193-006X}{\includegraphics[scale=0.06]{orcid.pdf}\hspace{1mm}Daragh Byrne} \\
  Carnegie Mellon University\\
  Pittsburgh, Pennsylvania, USA \\
  \texttt{daraghb@andrew.cmu.edu} \\
  \And
  \href{https://orcid.org/0000-0002-6839-796X}{\includegraphics[scale=0.06]{orcid.pdf}\hspace{1mm}Dina El-Zanfaly} \\
  Carnegie Mellon University\\
  Pittsburgh, Pennsylvania, USA \\
  \texttt{delzanfa@andrew.cmu.edu} \\
}

\hypersetup{
  pdftitle={WeldAR: Augmenting Live Hands-On Training with In-Situ Guidance for Novice Learners},
  pdfauthor={Chuhan (Franklin) Xu, Lia Sparingga Purnamasari, Zhenfang Chen, Daragh Byrne, Dina El-Zanfaly},
  pdfkeywords={AR In-the-Wild, Extended Reality, Augmented Reality, Tacit Knowledge, Physical Skill Learning, Embodied Interactions},
}

\begin{document}
\maketitle

\begin{center}
\small
Accepted to CHI 2026, Barcelona, Spain.\\
This is an author-created version of the work.\\
The Version of Record is published in the ACM Digital Library.\\
DOI: \url{https://doi.org/10.1145/3772318.3791913}
 
\end{center}

\begin{abstract}
 {Extended Reality (XR) systems for physical skill training have largely emphasized simulation rather than real-time in-situ instruction. We present WeldAR, an Augmented Reality (AR) system with five learning modules that overlays real-time guidance during live welding using a headset integrated into a welding helmet and a torch attachment. We conducted an in-situ within-subjects study with 24 novices, comparing AR guidance to video instruction for live welding across practice and unassisted tests. AR improved performance in \revised{both assisted practice and unassisted tests, primarily driven by gains in travel speed and work angle.}  By offering real-time feedback on four performance measures, AR supported novices in carrying embodied knowledge into independent tasks. Our contributions include: (1) WeldAR for in-situ physical skill training; (2) \revised{empirical evidence that AR enhances composite welding performance and key physical skills;} and (3) implications for the development of AR systems that support in-situ, embodied skill training in welding and related trades.}
\end{abstract}

\keywords{AR In-the-Wild \and Extended Reality \and Augmented Reality \and Tacit Knowledge \and Physical Skill Learning \and Embodied Interactions}

\begin{figure}[!htbp]
\centering
\includegraphics[width=\textwidth]{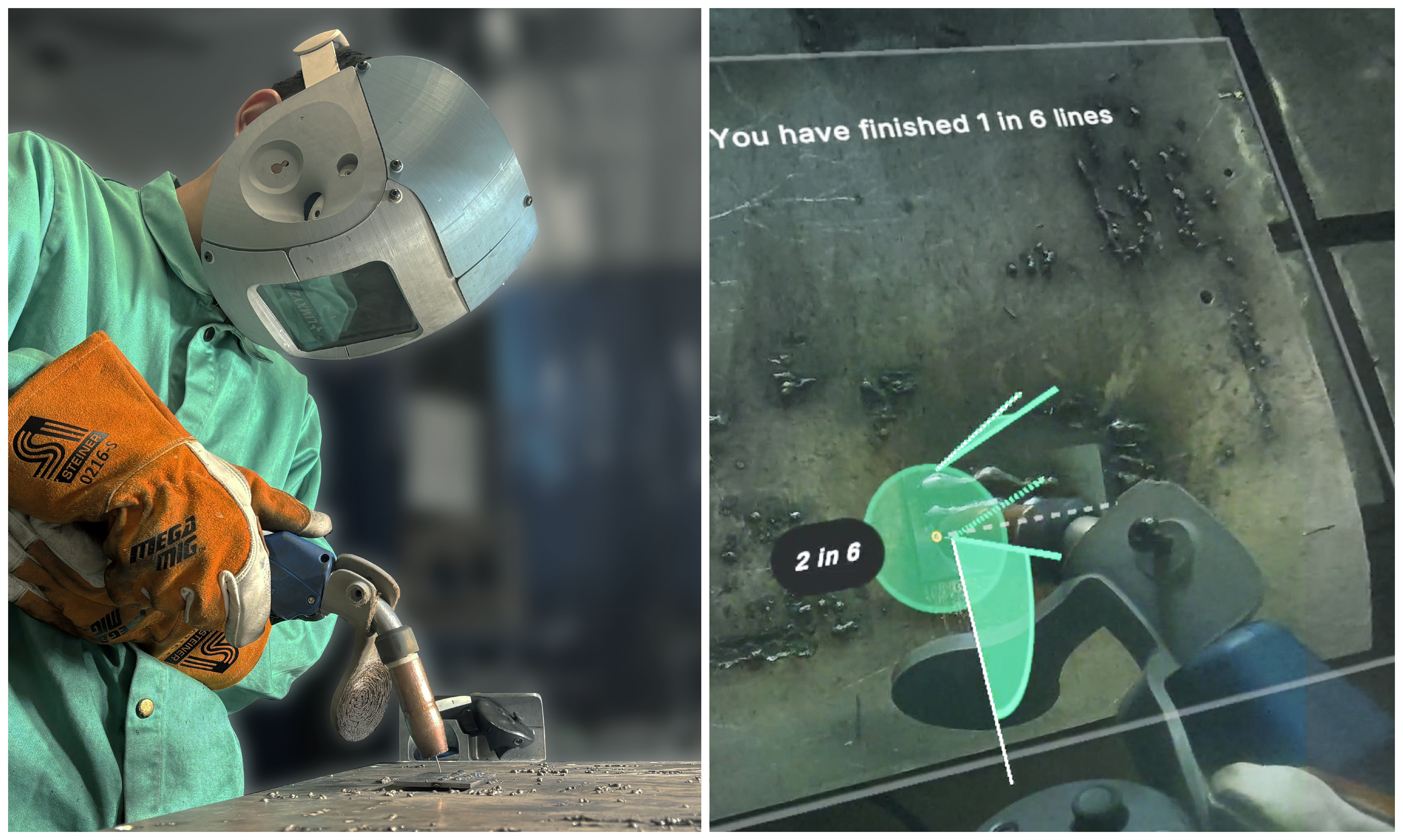}
\caption{Left: A student performs live welding with a customized helmet integrating a Meta Quest 3; WeldAR provides in-helmet guidance. Right: WeldAR’s interface displays real-time feedback during live welding on travel speed, standoff (CTWD), and work/travel angles to suggest technique adjustments.}
\label{fig:teaser}
\end{figure}

\section{Introduction}
Acquiring and performing physical skills in vocational trades, such as welding, depends on in-situ apprenticeship. In this intensive model, instructors provide real-time feedback and correction during repeated workshop practice, and novices progressively internalize tacit, embodied know-how that is best communicated through demonstration and situated feedback \citep{Ipsita2022-lq}. In welding training specifically, the learning environment presents unique instructional challenges. Most practice takes place in individual welding booths, with each booth typically accommodating a single learner. Within the booth, glare and personal protective equipment (PPE) limit the instructor's line of sight, while noise complicates real-time verbal coaching during welding. Consequently, instructors often rely on inspecting practice welds performed by novice students post hoc, rather than providing continuous, real-time observation and correction \citep{Asplund2020LessonsFT}. As a result of these limitations on continuous instruction, observation, and real-time correction at each welding booth, one-to-one instruction is challenging to scale to larger cohorts \citep{Lassiter2023WeldingIP}. 

These instructional challenges have led instructors to explore supplementary training methods. While video tutorials are often used to supplement limited one-to-one instruction in physical skill training, they provide only static instruction and typically lack the real-time feedback necessary for early skill development \citep{Fidalgo2023ASO}. Moreover, video training presents unique challenges in welding contexts—the high arc brightness, glare, and spatter make it difficult to capture and consistently show the weld puddle with conventional cameras. Rather than relying on conventional video approaches, live extended reality (XR) augmentation can deliver guidance at the welder's line of sight via visual overlays on the welding workbench while welding. Because these visual overlays are synthesized from tracking data and presented at the welder's line of sight through an XR headset, they remain visible when visual conditions are impaired due to glare and spatter, enabling personalized, real-time feedback at each booth.

Building on this rationale, evidence across training domains shows that XR, encompassing virtual reality (VR), augmented reality (AR), and mixed reality (MR), can scale instruction and deliver in-situ, real-time feedback \citep{Bschel2021MIRIAAM, Zenner2021HaRTT}. For example, in medical education, XR-based simulations help students practice interpreting subtle sensory cues and coordinating motor actions, leading to improvements in early surgical performance and reduced reliance on direct instructor oversight \citep{wang_critical_2025,chan_use_2026,stone_autonomous_2025}. In manufacturing and machining training, XR systems scaffold embodied procedural skills, allowing learners to gradually build competency in complex hand-eye and tool-material coordination before transitioning to full-scale practice \citep{Huang2021AdapTutAR}. Most welding-focused XR training systems operate in simulated environments \citep{Chan2022VRAA}. These VR- and AR-based welding simulators have shown promise for accelerating early learning and improving novice performance \citep{AkundiAugmentedRI,Byrd2015TheUO}.

However, simulations cannot fully reproduce the situated nature of live welding, including arc brightness, sparks, heat, material distortion, and auditory and haptic feedback. As a result, the skills acquired in simulation often require recalibration when transferred to live welding tasks, resulting in a performance gap. This gap motivates the study of how AR can directly augment live welding, support in-situ training, and improve the transfer of skills to professional practice. 

Accordingly, this paper focuses on live, in-situ AR instruction for MIG (Metal Inert Gas) welding, a beginner-friendly welding type widely used in educational training settings, such as an elective course offered at our institution. We developed WeldAR, an AR system designed for novice welders that augments live welding training. WeldAR supports basic welding skills, including parameters such as contact to work distance (CTWD), travel angle, work angle, and travel speed, through real-time sensing and feedback (see Figure~\ref{fig:teaser}). Learning these parameters is essential for mastering the basic skills of welding. The system combines a video see-through head-mounted display (HMD) with a custom 3D-printed helmet that includes a welding auto-darkening screen. This configuration manages the intense glare and arc brightness typical of welding while enabling the projection of low-latency visual overlays. Using WeldAR as a research probe, we investigate the following three questions:

\begin{itemize}
    \item \textbf{RQ1:} How does augmenting live welding with AR guidance affect novice performance on predefined indicators of physical skill development, such as travel speed variance, CTWD error, and work/travel angle deviation?
    \item \textbf{RQ2:} To what extent does AR guidance support the transfer of physical skills to independent, unassisted (no real-time feedback provided) welding tasks?
    \item \textbf{RQ3:} What ergonomic, usability, and experiential factors shape the effectiveness of AR systems in live welding training?
\end{itemize}

To answer these questions, we conducted a within-subjects crossover study with 24 novice participants who performed welding tasks under two instructional conditions: AR guidance and video-based instruction (as a surrogate for an instructor). The study included assisted (with displayed real-time AR feedback) trials and unassisted (no real-time feedback provided) transfer tests. Alongside performance measures, we collected surveys and interviews to examine usability, workload, and learner perceptions of AR support in real workshop settings.

Through this study, we make four contributions:

\begin{enumerate}
    \item WeldAR, an AR system for live welding training, including hardware configuration, sensing pipeline, and instructional modules.
    \item A within-subjects user study with 24 novices comparing AR guidance to video-based instruction, measuring both assisted and unassisted performance.
    \item Empirical insights into the usability and experiential factors that influence the effectiveness of AR in hands-on vocational training.
    \item Implications for the development of AR systems that support in-situ, embodied skill training in welding and related trades.
\end{enumerate}

In this paper, we first review related work on XR for physical skill training, augmenting live hands-on training, and welding training. We then present WeldAR's hardware, sensing pipeline, and instructional overlays. Next, we describe the user study design and measures, followed by quantitative and qualitative findings. We conclude with a discussion of implications and limitations, and a conclusion summarizing our work.

\section{Related Work}

\subsection{XR for Physical Skill Learning}

Extended Reality (XR) has emerged as a promising approach to support embodied learning across various domains. XR environments offer affordances such as immersion, visuo-motor feedback \citep{Yu2024DesignSO}, and the ability to simulate complex scenarios that would otherwise be costly, unsafe, or resource-intensive to reproduce \citep{jensen2018vr}. XR has been used to support physical skill learning across domains such as medical training, industrial assembly, and craft practice by providing immersive guidance and real-time feedback \citep{Uhl2023,xu2021hmd,Huang2021AdapTutAR,Ji2025,wang_impact_2024,OrigamiSensei,AugmentingWeldingTraining}.

Welding has also adopted XR-based training tools. Commercial VR simulators such as Soldamatic \citep{Seabery2024} and Miller AugmentedArc \citep{Miller2023} allow learners to practice welding techniques in safe, low-cost virtual environments. \revised{Research and commercial platforms such as VRWELDLearner \citep{Ipsita2022-lq} and WeldVR \citep{weldvrWeldingSimulator}  use HMD-based VR environments and instrumented torches to train and assess foundational welding parameters.} These systems typically replicate basic welding tasks (e.g., bead placement on flat or groove joints) while tracking parameters like travel speed, arc length, work angle, and travel angle. Research has shown that XR welding simulators improve welding performance outcomes including weld quality scores, welding parameter scores, and dexterous abilities \citep{preston_byrd_dexterity_2018,huang_research_2020,liang_simple_2014,wells_effect_2020}. These improvements are especially pronounced in the early stages of learning, where simulators increase student engagement, reduce material waste and training costs, and allow novices to develop basic torch handling skills before moving to the workshop \citep{shankhwar_interactive_2022,shankhwar_visuo-haptic_2022,chung_research_2020,rodriguez-martin_learning_2019}.

However, a systematic review \citep{Heibel2023VirtualRI} highlights that most XR training systems excel at face and concurrent validity but often lack predictive validity, or evidence that skills learned in XR transfer seamlessly to real-world practice. This reflects a broader trend in XR research: systems are frequently designed for simulated environments, where sensory fidelity and environmental constraints are simplified. \revised{In welding, most XR systems, primarily VR-based, prioritize simulation fidelity (e.g., visual realism and scoring) over the sensory and embodied constraints of live practice \citep{Chan2022VRAA}. This matters for transfer because welding is learned through continuous adjustments of CTWD, torch angles, and travel speed based on weld-pool visibility, arc sound, heat, and glare. When these cues are simplified or absent, as in many simulators, learners may achieve high simulator scores yet still struggle to recalibrate control under live welding conditions.}

\revised{As a result, learners often experience a gap when transitioning from simulation to live welding, requiring recalibration under real-world conditions. This motivates examining XR guidance that supports in-situ practice welding in high-risk settings.}

\begin{figure}[!htbp]
\centering
\includegraphics[width=\columnwidth, alt={The left panel shows an augmented reality welding interface with green guidance overlays around the torch tip, a dotted white guideline indicating torch alignment, and a progress widget labeled “2 in 6.” The right panel shows a Welding Summary screen with four metrics - CTWD, travel angle, work angle, and travel speed. Each displayed as a bar chart with target ranges, percentages, and labels indicating time spent within or outside the acceptable range.}]{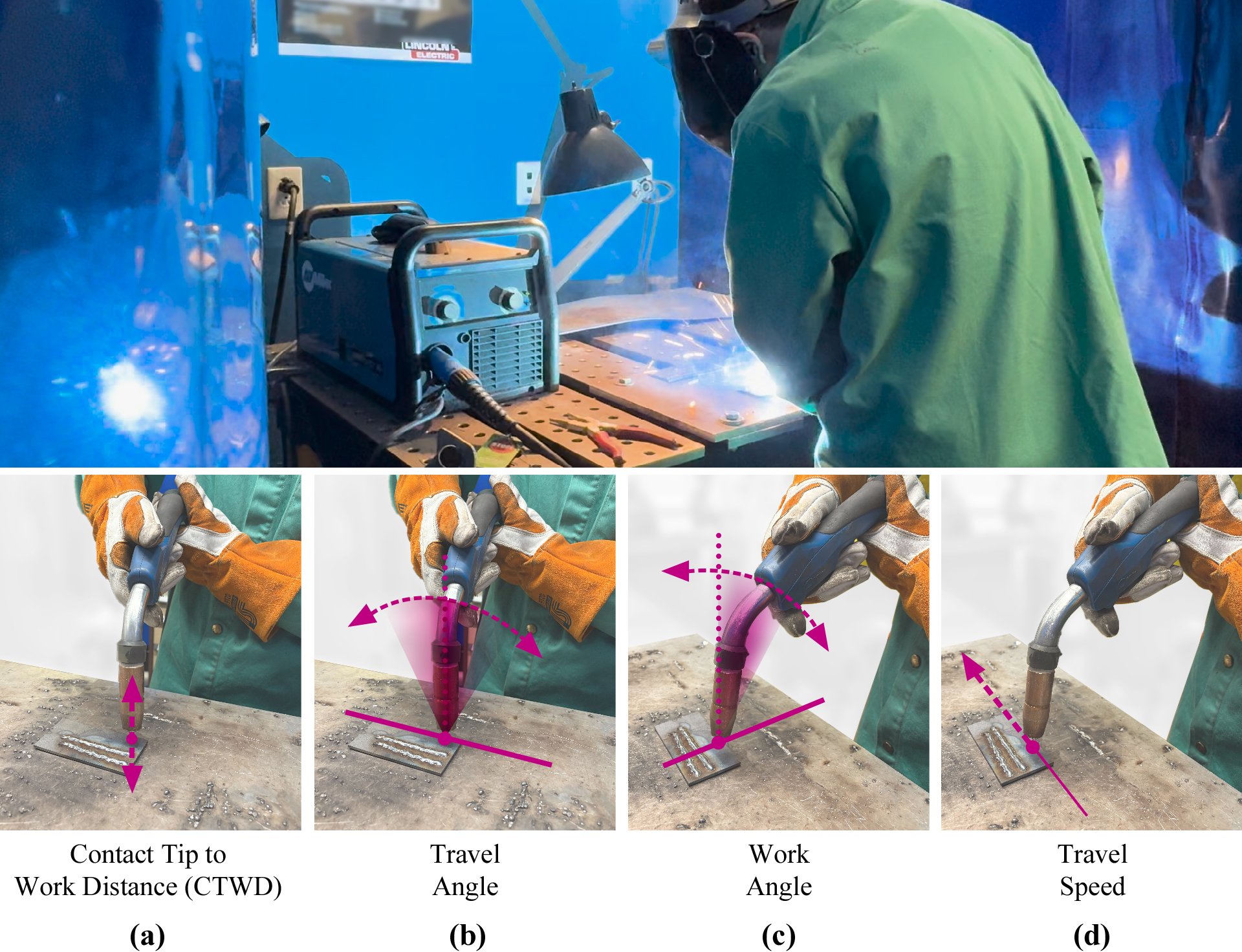}
\caption{A novice student is learning how to weld in the welding booth by learning fundamental skill parameters, including (a) Contact Tip to Work Distance (CTWD), (b) Travel angle, (c) Work angle, and (d) Travel speed.}
\label{fig:weld_training}
\end{figure}

\begin{figure}[!htbp]
\centering
\includegraphics[width=\columnwidth, alt={Two metal welding vouchers with welded lines on them. The one on the left shows four uneven welded beads, and the Voucher on the right shows three even welded beads}]{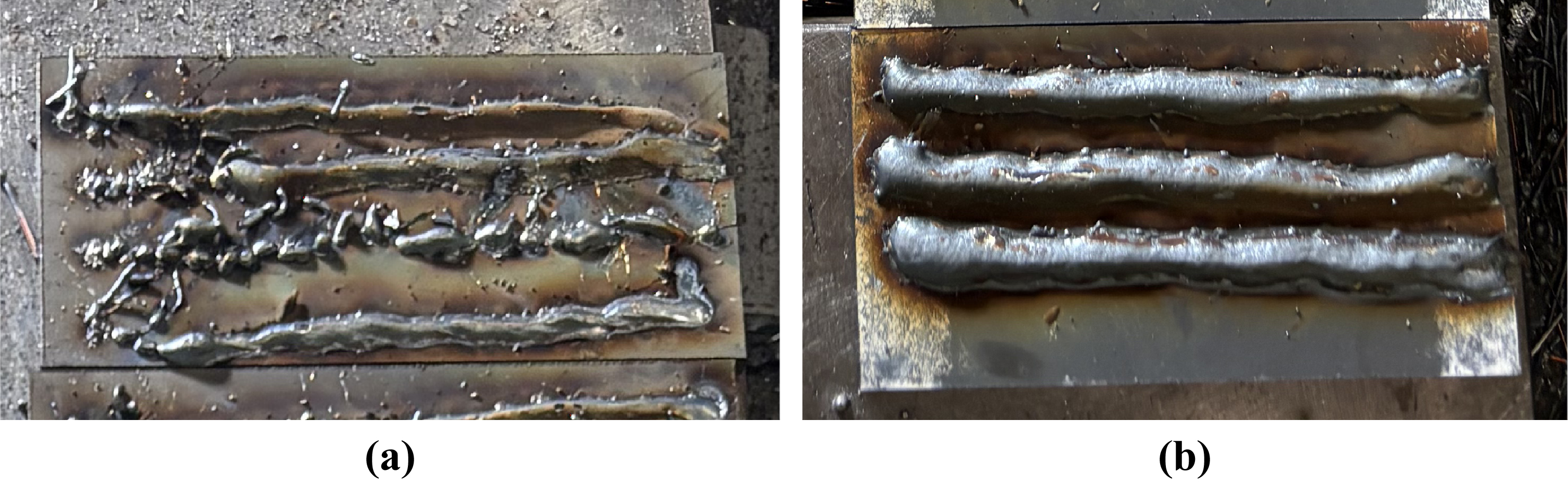}
\caption{Welding coupons that students use to train on welding. \textbf{(a)} shows broken weld lines and \textbf{(b)} shows ideal welded lines. Image \textbf{(b)} indicates that the novice student stayed within the acceptable range of the welding parameters, including CTWD, Travel Speed, Travel Angle, and Work Angle. While (a) indicates that the novice student performed outside the acceptable range.}
\label{fig:voucher}
\end{figure}

\subsection{Augmenting Live Hands-On Training}

A growing body of work explores augmenting live skill practice. In musical training, ImproVISe \citep{Deja2024TeachMH} projected harmonic and rhythmic guidance directly onto the keyboard, illustrating how hardware projection and carefully timed visual content can scaffold improvisation when feedback is aligned with musical context and learner confidence. In motor-skill training, AR Hoop \citep{Turakhia2025} dynamically adjusted the size of a virtual target during ball-throwing, showing how adaptive hardware-software integration and multimodal cues such as depth, sound, and tactile feedback are critical for effective motor learning. In exercise, research \citep{Adiwangsa2025} explored overlaying AR guidance onto everyday objects, showing how lightweight hardware and context-sensitive feedback can tailor learning content to individual needs while balancing structure with novelty.

In craft training, systems have used haptics and sensing to surface otherwise hidden dynamics (e.g., EmbodyCraft \citep{Zhu2025}) and combined holographic gesture guidance with real-time material feedback (e.g., wheel-throwing \citep{Ji2025,gao_digiclay_2018}). BRICKxAR \citep{Yan2021} highlights the need to augment sub-millimeter marker-based registration, realistic object occlusion, and hand occlusion, enabling users to “grasp” virtual bricks in ways that preserve spatial coherence.

Recent XR welding training systems have begun to integrate live welding. XR Bot Trainer \citep{Lee2023-ca} augments welding actions with real-time overlays. \revised{It relies on external cameras and a fixed-view VR display that shows a reconstructed scene instead of the physical environment, which can introduce latency and restrict viewpoint flexibility. Additionally}, its instructional content is limited to flat-plate seam tracking and does not address the wider set of welding parameters that novices must master.

\revised{Prior work demonstrates the promise of live augmentation, but leaves a key gap for welding: systems either focus on VR simulation or provide limited live guidance under restricted viewpoints and task settings. There is limited prior work on low-latency, head-tracked AR guidance during live welding that supports the full set of novice-relevant motor skills.}

\subsection{Welding Education and Training Needs}
Welding is often taught through apprenticeships and vocational programs. Instructors supervise students during training, observing their technique, pointing out errors, and demonstrating the correct welding posture, speed, distance, or angle \citep{Asplund2020LessonsFT}. Students improve through repeated practice under supervision to develop the perceptual sensitivity and motor coordination required for proficiency. 

Instruction for beginners focuses on foundational welding parameters that directly affect bead quality. Staying within acceptable ranges improves weld bead quality, compared to performing outside the acceptable range (Figure~\ref{fig:weld_training}). These parameters include CTWD, travel speed, work angle, and travel angle. The acceptable ranges of each of these parameters depend on the welding machines used, the feed rate, and the thickness of the metal piece, called a coupon (see Figure~\ref{fig:voucher}), being welded \citep{miller2024migparameters}.

Because novices have not yet developed the perceptual and motor skills needed to monitor and adjust these parameters independently, they rely on instructors for feedback. However, this level of repeated, individualized oversight is increasingly difficult to sustain due to a shortage of qualified instructors, creating a significant challenge for vocational training programs \citep{lassiter2023ai_welding}.

\section{WeldAR}

 \begin{figure*}[!htbp]
\centering
\includegraphics[width=0.75\linewidth, alt={Three images on the right above each other showing custom hardware components of the AR welding system. The first image displays a Meta Quest 3 headset integrated into a custom 3D-printed welding helmet. The second image shows a 3D-printed controller bracket attached to a welding torch. The third image shows a compact 3D-printed calibration station mounted on the welding bench, used to align the torch tip before practice. The diagram on the left shows the system diagram of how all these components are connected}]{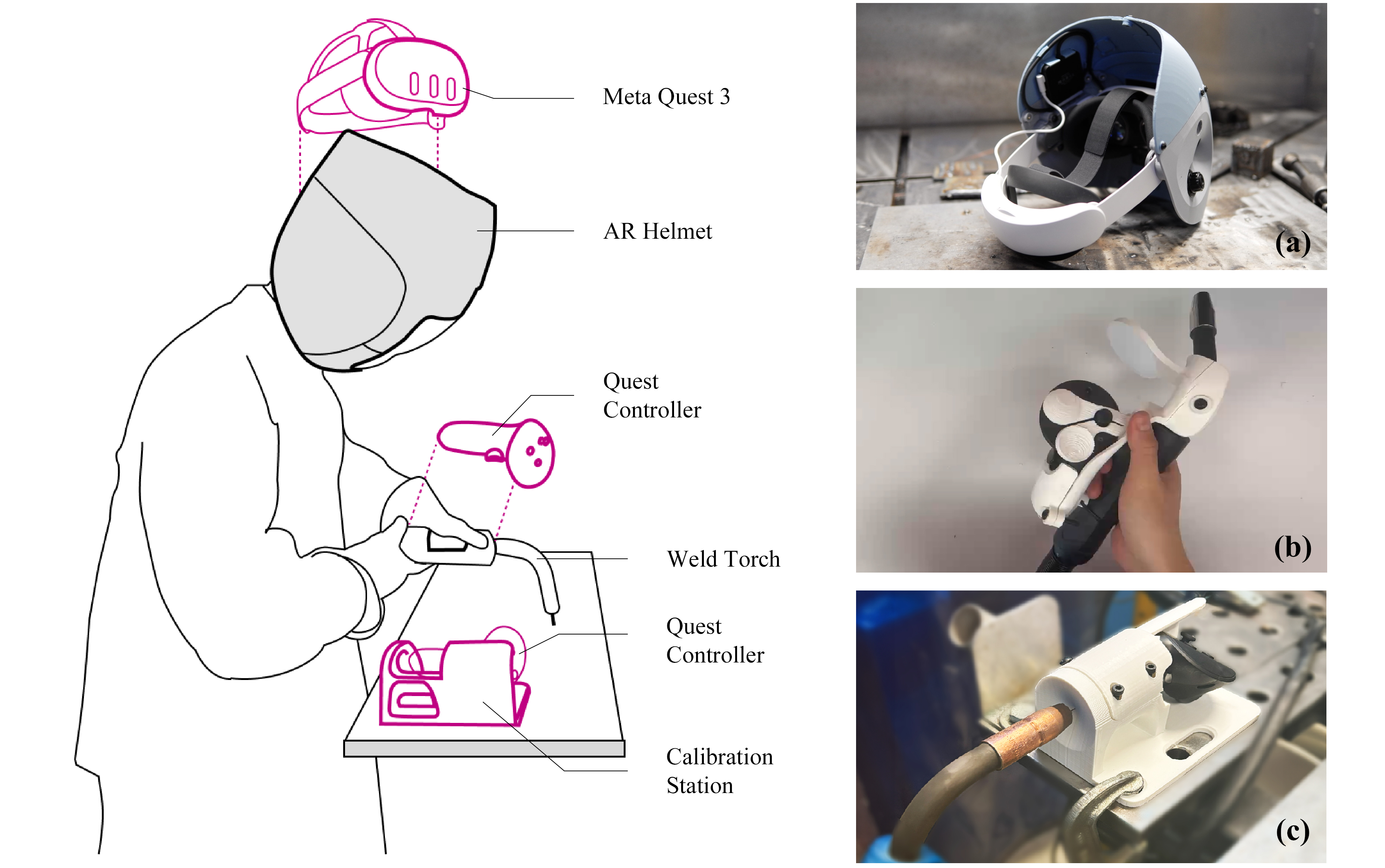}
 \caption{The images on the right show the 3D-printed hardware components of WeldAR: \textbf{(a)} Customized 3D-printed welding helmet equipped with a Meta Quest 3 headset \textbf{(b)} Customized 3D-printed controller bracket attached to the welding torch \textbf{(c)} Customized 3D-printed calibration station attached to the workbench. The systems diagram on the left shows a student learning MIG welding with the setup, and how all these components are connected.}
\label{fig:hardware}
 \end{figure*}
\revised{Building on prior XR welding limitations, WeldAR introduces a multi-parameter, in-situ AR approach designed for the embodied and sensory demands of live MIG welding.}

\subsection{\textbf{System Description}}
WeldAR is an augmented-reality (AR) system that delivers real-time visual feedback and stepwise in-helmet instruction for live MIG welding. The design reflects a 26-month collaboration with a non-profit welding training organization for teenagers and young adults. This includes four co-design workshops with instructors and novices to develop and evaluate the system iteratively in-situ.

The system is designed to be used during live welding. WeldAR tracks torch pose and motion, computes four core skill parameters: CTWD, work angle, travel angle, and travel speed. It provides real-time visual and verbal feedback, along with post-hoc summaries. The hardware supports unobstructed, safe operation in real welding settings, while the software captures, processes, and presents motion data to enable accurate feedback and skill acquisition (Figure~\ref{fig:hardware}).

The system comprises a Meta Quest 3 headset integrated into a custom 3D-printed welding helmet, a torch-mounted instrumented attachment, and a bench-mounted calibration station. Built into the headset, a scaffolded curriculum delivers four single-parameter modules followed by a compound module that integrates all parameters for flat-plate practice:

\begin{itemize}
    \item \textbf{Contact Tip to Work Distance (CTWD)}: maintaining a certain distance between the tip of the welding torch and the coupon;
    \item \textbf{Work angle}: the lead angle of the torch relative to the weld surface normal;
    \item \textbf{Travel angle}: holding the torch relative to the direction of motion along the weld line;
    \item \textbf{Travel speed}: the torch movement speed along the weld line. 
    \item \textbf{Combination}: a compound pass that integrates all four parameters for simultaneous practice and evaluation.
\end{itemize}

These learning modules provide the instructions needed for learners to develop the physical skills required for learning early-stage live welding.

\subsubsection{\textbf{Hardware Development}}

\begin{figure}[!htbp]
\centering
\includegraphics[width=0.75\columnwidth, alt={Diagram illustrating two workflows in the data collection pipeline. Workflow 1 (sound trigger): welding torch (A) produces sound, captured by microphone (B), transmitted to laptop (C), and then signals sent to the headset (D). Workflow 2 (performance data collection): Motion data is captured by controller (E), transmitted to headset (D), and sent to laptop (C). Additional system components include work light (F) for stable tracking and calibration station (G) for alignment.}]{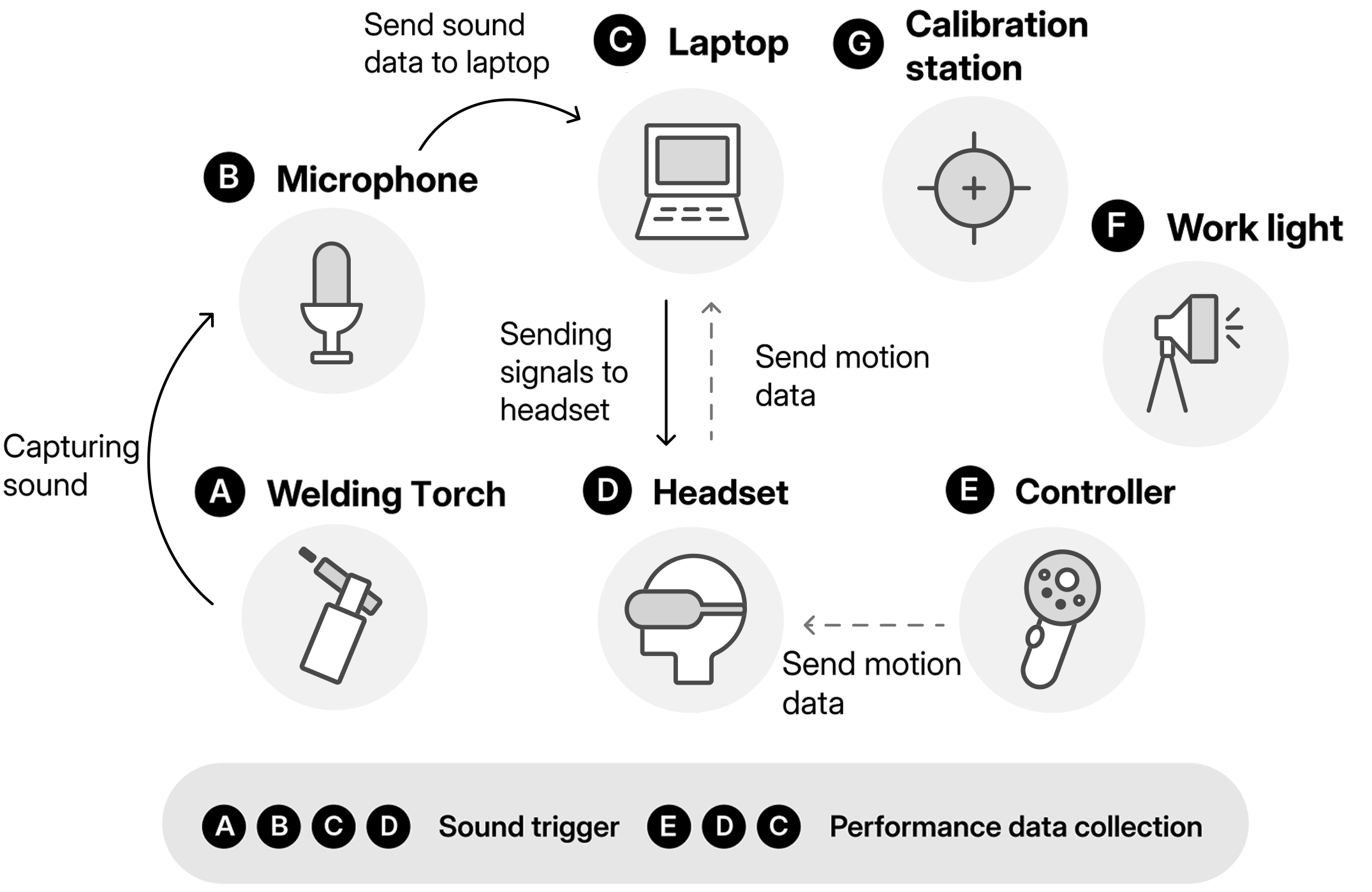}
\caption{System workflow diagram showing sound trigger and performance data collection.
}
\label{fig:system_workflow}
\end{figure}

\begin{figure*}[!htbp]
\centering
\includegraphics[width=0.68\linewidth, alt={Diagram illustrating the hardware setup for an AR-assisted welding system. The top section shows the welding torch with a mounted right-hand controller, microphone, trigger lever, buttons, protection cover, and controller holder. The lower section displays the calibration station with a left-hand controller, calibration point, and different mounting options. On the right, the figure shows a customized welding helmet with a 3D-printed mount, auto-darkening screen, and a Meta Quest 3 headset inside. Labels identify each major component.}]{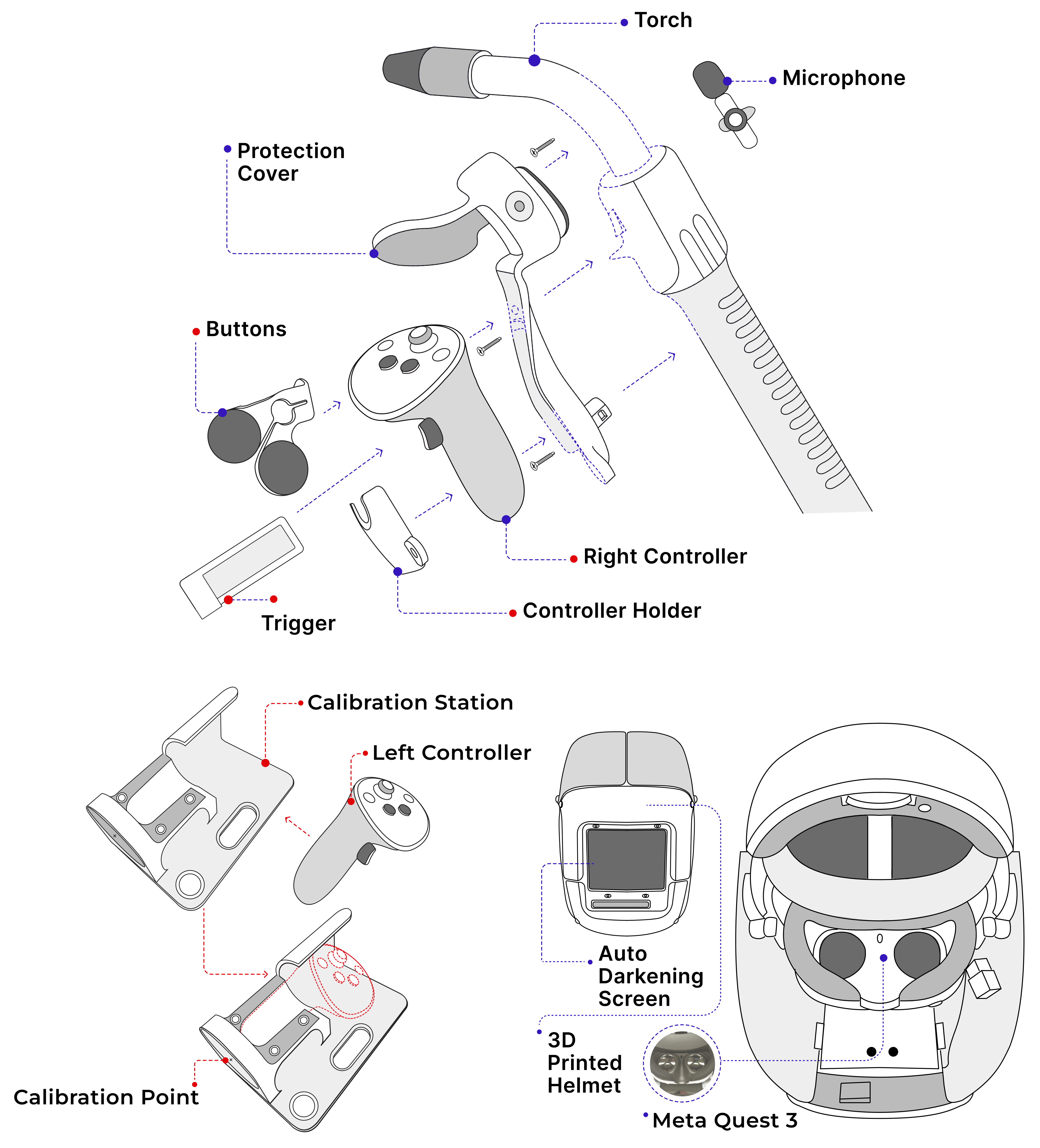}
\caption{Hardware diagram showing the customized helmet, torch, calibration station, and weld trigger components.}
\label{fig:hardware_workflow}
\end{figure*}

\paragraph{Helmet Design.} To support safe AR use during welding, we designed a custom welding helmet that integrates the Meta Quest 3 as the primary AR interface. The front section of the helmet accommodates an auto-darkening screen and exterior shield glass, both standard safety features. \revised{Its modular design includes accessible mounting points (Figure~\ref{fig:hardware_workflow}) to support local 3D printing and quick sensor replacement.} For broader adoption, we chose the low-cost Meta Quest 3. We paired it with Quest Pro controllers for torch tracking instead of the Quest 3 controllers to maintain tracking accuracy when the auto-darkening screen is in use.

\paragraph{Torch Design.} \revised{We developed a custom 3D-printed attachment to mount a Quest Pro controller on a standard welding torch, enabling motion tracking while preserving realistic handling.} A protective shield reduces spark exposure and improves equipment safety for the Quest Pro controller during live welding. While early prototypes aimed for cross-model compatibility, the final version was tailored to the partner organization’s equipment, and both right- and left-handed variants were produced for training settings.

\paragraph{Calibration Station.} To anchor the AR world coordinate frame, we designed a calibration station housing a Quest Touch Pro controller. Clamped to the workbench, it aligns the virtual and physical systems and supports torch calibration via a precision rear groove (Figure~\ref{fig:hardware}c). Multiple mounting options, including screw holes and clamp slots, make it adaptable to different welding benches. At the start of each session, the headset establishes a spatial anchor at the calibration station; if tracking is lost, the system enables the user to relocalize the torch to this anchor to restore alignment and correct drift.

\paragraph{Welding Trigger Components.} To ensure accurate system activation, we implemented two triggering methods based on the welding torch model. 

\begin{enumerate}
    \item Mechanical: a lever attached to the torch trigger that presses the controller button when activated. 
    \item Acoustic: an independent Bluetooth lavalier microphone (with noise suppression settings turned off) used to trigger the system when it detects weld sounds above a volume threshold. 
\end{enumerate}

Each method operates independently and shows low latency, helping to minimize disruption during welding and reduce additional user actions. Together, these components underpin two core workflows: synchronizing AR guidance with welding onset, and continuously capturing motion data for feedback and analysis (see Figure~\ref{fig:system_workflow}).

\subsubsection{\textbf{Software Development}}
\begin{figure}[!htbp]
\centering
\includegraphics[width=\columnwidth, alt={The left panel shows an augmented reality welding interface with green guidance overlays around the torch tip, a dotted white guideline indicating torch alignment, and a progress widget labeled “2 in 6.” The right panel shows a Welding Summary screen with four metrics—distance, travel angle, work angle, and speed—each displayed as a bar chart with target ranges, percentages, and labels indicating time spent within or outside the acceptable range.}]{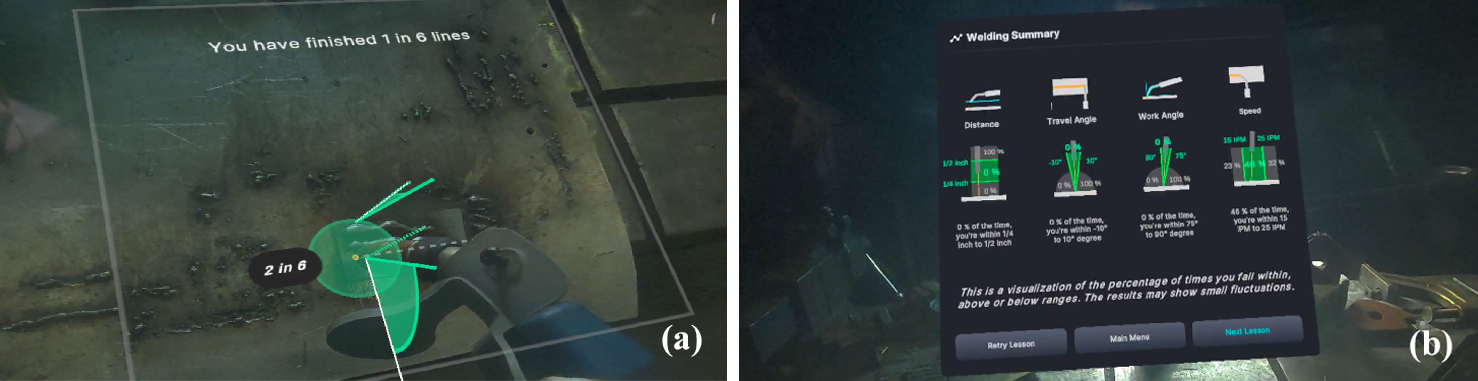}
\caption{\textbf{(a)} WeldAR's UI widgets and a dotted white line attached to the torch tip for real-time guidance and feedback during live welding. \textbf{(b)} A summary at the end of each module provides overall performance feedback.}
\label{fig:software}
\end{figure}

\begin{figure}[!htbp]
\centering
\includegraphics[width=\columnwidth, alt={Three images showing the recalibration process in the AR welding system. In the first two panels, a glowing magenta dot appears on the metal weld plate, and the user aligns the physical torch tip with the dot. In the third panel, the virtual overlay reappears, showing a glowing green torch tip, guideline arcs, and a progress indicator labeled “3 in 4,” confirming that the virtual torch has been realigned with the physical tip before starting the weld line.}]{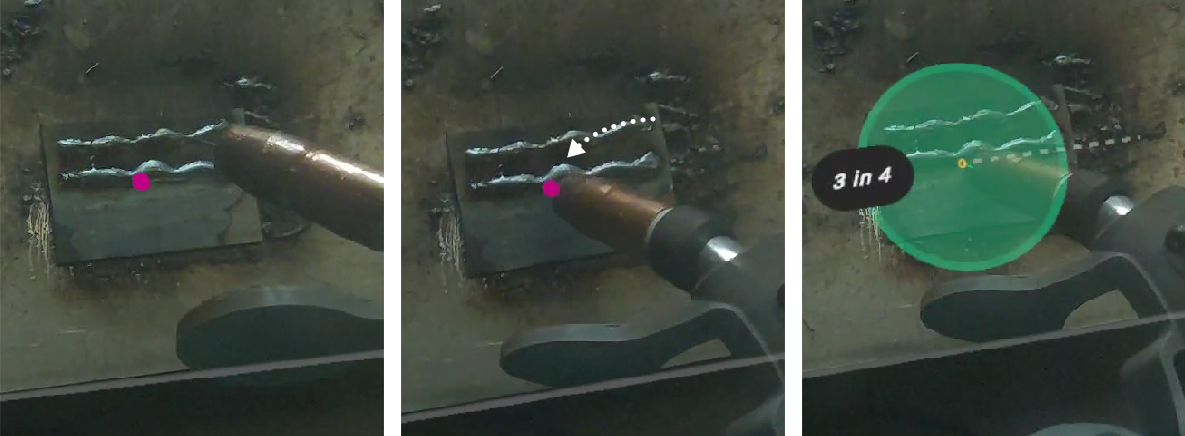}
\caption{Torch recalibration per line. Before the start of each line, a glowing dot appears on the coupon. The user touches the dot with the torch tip, and realigns the virtual tip to it.}
\label{fig:calibration}
\end{figure}

\paragraph{Digital Assets and Calibration.} \revised{We align digital twins of the torch and workbench to derive four skill parameters: CTWD, work/travel angles, and travel speed. To mitigate tracking errors \citep{10.1145/3463914.3463921, act12060257} and drift from headset removal, we implemented two calibration routines: a general system realignment and a rapid per-line reset in which the user touches a projected target with the torch tip (Figure~\ref{fig:calibration}).}

\begin{figure}[!htbp]
\centering
\includegraphics[width=\columnwidth, alt={A set of eight AR welding training screenshots illustrating live UI feedback for four welding skills. For CTWD (contact tip to work distance), the correct range is shown with a green circle, while an out-of-range state shows a red overlay labeled “Too far from table.” For travel angle, the correct orientation is indicated with a green arrow overlay, while incorrect alignment is shown with a red overlay and guiding arrows. For work angle, a green wedge indicates within range, and a red overlay with directional arrows indicates deviation. For travel speed, a green bar shows correct movement, while a red bar with a label such as “Too fast” signals that the speed is outside the acceptable range.}]{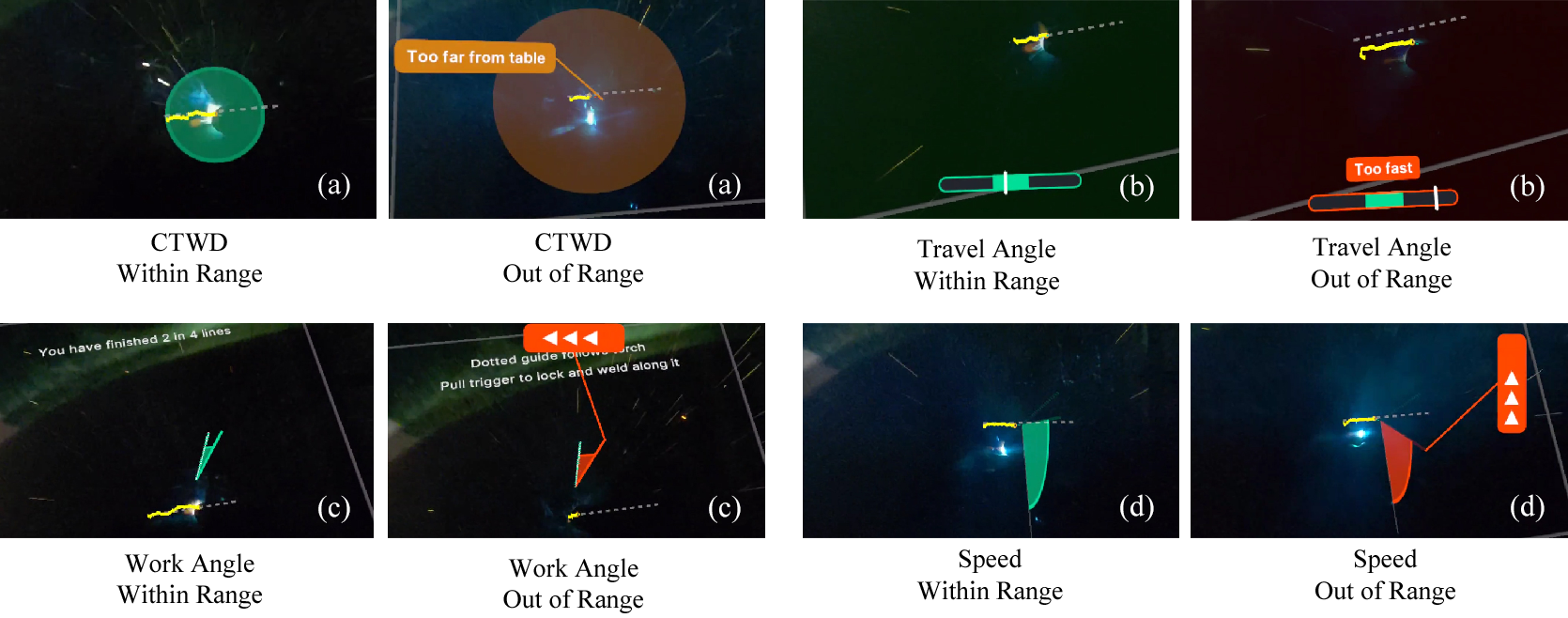}
 \caption{Live UI feedback for each skill parameter during welding.\textbf{ (a)} CTWD is shown with a green circle when within range and a red overlay with “Too far from table” or “Too close from table” labels when out of range. \textbf{(b) }Travel angle feedback appears as a green overlay when correct and a red overlay with left/right arrows when misaligned. \textbf{(c)} Work angle is indicated with a green overlay when within range and a red overlay with forward/backward arrows when out of range. \textbf{(d)} Travel speed is displayed with a green bar when appropriate and a red bar labeled “Too fast” or “Too slow” when outside the target range.}
\label{fig:comparisons}
 \end{figure}

\paragraph{User Interface Design.} 
\revised{To balance information proximity with minimal occlusion, we designed a torch-mounted interface based on pilot studies' (workshops conducted at our partner site) findings that distant overlays increase cognitive load while direct placement obstructs the weld pool. Adhering to spatial computing safety guidelines \citep{metaConsiderations}, semi-opaque widgets were positioned at the periphery to keep the central view clear.} 
Feedback utilizes accessible color coding (green/red/blue-orange) to indicate status, appearing primarily when values exit target ranges to minimize distraction (Figure~\ref{fig:comparisons}). 
\revised{Additionally, a projected dotted path locks onto the work surface upon activation to guide linear torch movement.}

\paragraph{Lessons.} To support learning, we introduced the four skill parameters through sequential modules or lessons. The lesson plan introduces tracked variables step by step, so beginners would not be overwhelmed \citep{klotzbier2025scaffolding}. At the end of the lessons, a combination-task module brings the four parameters together, encouraging learners to integrate the skills and practice. This scaffolded design enables students to build confidence by practicing one aspect at a time before combining them.

\paragraph{Performance Feedback.} \revised{Post-lesson summaries evaluate proficiency based on the percentage of time spent within target ranges \citep{miller2012om238, miller2024migparameters}. Students must meet a specific threshold to advance (Figure~\ref{fig:software}). Supplementary metrics include “smoothness” (variance) and “accuracy” (path deviation), alongside a replay function. Finally, an optional unassisted mode disables feedback for benchmarking.}

\subsubsection{\revised{\textbf{Distinctions from Prior Welding Training Platforms}}}

\revised{WeldAR is designed for in-situ, live MIG welding instruction, rather than pre-weld simulation. This design choice targets the sensory and embodied constraints of real welding practice (e.g., heat, glare, noise) and supports novices in coordinating four core motor-skill parameters through head-tracked, low-latency pass-through AR cues aligned with real welding torch motion. In contrast, XR Bot Trainer \citep{Lee2023-ca} focuses on TIG welding and provides guidance for a single cue (path alignment/speed) via external cameras and a fixed, reconstructed VR view that can introduce higher latency and restrict viewpoint flexibility. Likewise, VR simulators such as WeldVR \citep{weldvrWeldingSimulator} and VRWELDLearner \citep{Ipsita2022-lq} are fully virtual, pre-weld systems that rely on VR controller-triggered virtual welding. Specifically, WeldVR operates without the physical fidelity of a real torch and provides no real-time cues for CTWD or angles. Conversely, VRWELDLearner does not scaffold learning into structured modules and does not report performance or evaluate transfer to live welding. Overall, WeldAR differs by delivering spatial guidance during actual welding tasks, directly supporting the motor control required for real torch movement and angle regulation.}

\subsection{\textbf{Data Formulation}}
\revised{We process controller tracking data at 90 Hz to compute four real-time skill parameters and generate feedback.}

\subsubsection{\textbf{Live Frame Collection}}

\revised{We calculated parameters based on the 6-DOF pose of the virtual torch relative to the calibrated reference grid. \textbf{CTWD} is computed as the perpendicular distance from the virtual torch tip to the grid plane, while \textbf{Travel} and \textbf{Work Angles} are derived from the angle between the torch’s forward vector and the vertical or horizontal axes relative to the welding direction. To determine \textbf{Travel Speed}, we smooth torch displacement using a 2D Kalman filter \citep{10.1145/3181673} to suppress hand jitter, then compute the average velocity projected onto the weld-direction unit vector over a sliding window, converting the result to inches per minute (IPM).}

\subsubsection{\textbf{Performance Feedback}}
\begin{figure}[!htbp]
\centering
\includegraphics[width=0.75\columnwidth, alt={A diagram showing the target ranges for four welding parameters. For CTWD (contact tip to work distance), the acceptable range is 6–15 mm above the surface. For travel angle, the correct range is within ±10° from the vertical centerline. For work angle, the acceptable range is 75–90° relative to the surface (within 15° of perpendicular). For travel speed, the range is between 15 and 25 inches per minute (IPM).}]{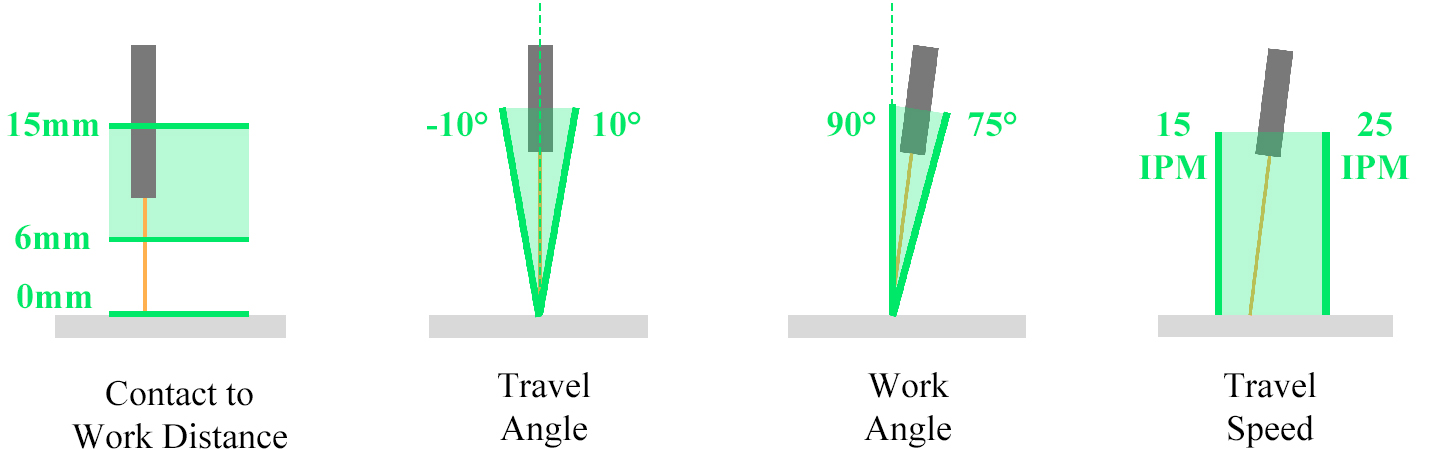}
 \caption{A diagram showing the acceptable range for each skill parameter, for CTWD it is between 6 mm and 15 mm, for travel angle the range is 10 degrees for both sides. For work angle, the range is 15 degrees from the perpendicular line. For speed, the range is 15-25 IPM.}
\label{fig:criteria}
 \end{figure}

\revised{We defined target ranges based on welding standards \citep{miller2012om238,miller2024migparameters} to drive real-time UI triggers (green: within acceptable range; red: out of range) and evaluation. The system targets include \textbf{CTWD $\in [6, 15]$ mm}, \textbf{Travel Angle $\in [-10, 10]^\circ$}, \textbf{Work Angle $\in [75, 90]^\circ$}, and \textbf{Travel Speed $\in [15, 25]$ IPM}.} These ranges directly influence real-time feedback: the UI color changes dynamically based on the current frame’s state, and feedback is provided whenever the user moves outside the acceptable range (see Figure~\ref{fig:criteria}). After each lesson, a performance evaluation is carried out by counting the number of frames that fall within, above, or below the target range, and dividing each count by the total number of frames to compute percentages. These percentages are then presented in the UI as a summary for each lesson.

\subsection{\textbf{Data validation for In-the-Wild Deployment}}

Ahead of our user study, we conducted validation studies to confirm system reliability in the welding booth. These tests focused on controller drift, as well as data accuracy and sound-trigger latency. We then implemented mitigation strategies based on the results. Detailed procedures are available in Appendix~\ref{app:system_validation}.

\paragraph{Drift and Mitigation. } Prior work has shown that Quest headsets and controllers exhibit minor jitter and drift in tracking \citep{10.1145/3463914.3463921, act12060257}. \revised{In our welding booth with the auto-darkening screen and live arc active, headset tracking remained stable; we only observed occasional controller drift, typically appearing as sudden CTWD/speed jumps while rotation data remained stable.} Across 40 trials of 30s duration each, about 7\% of frames were affected and 25\% of weld lines contained at least one drift event. Bootstrap simulations (10,000 samples) estimated worst-case probabilities of up to 50\% for four- and six-line sequences. To mitigate this, we: (1) provided per-line tap-to-recalibrate at the calibration station; (2) improved task lighting to stabilize tracking; (3) surfaced drift in the UI during lessons; and (4) flagged/excluded drift-affected frames/lines in analysis while increasing practice/test counts to maintain statistical power.

\paragraph{Accuracy Checks.} To verify that key skill parameters could be reliably captured in live welding settings, we ran bench tests with jigs and measured wire lengths. \revised{Results showed: travel angle error M = –0.44°, SD = 0.40°; work angle error M = 0.99°, SD = 0.78°; CTWD error M = 2.59 mm, SD = 1.75 mm; and speed deviations for 3–7 inch lines between 0.56 and 1.01 IPM (SD = 0.35–0.72).} Angles and speed were more stable than displacement-based measures, while CTWD errors were larger but remained within usable bounds for feedback.

\paragraph{Audio Timing and Alignment.} To address the latency introduced by Bluetooth microphone transmission, we measured the detection delay of the sound trigger, and observed an offset on the order of milliseconds. Based on these results, we chose a low-latency buffer and shifted analysis windows by \~280 ms (20 frames at 90 Hz) to align sound triggers with motion logs.

\section{User Study}

The goal of this study was to evaluate how augmenting live welding practice with AR guidance supports novice performance, skill transfer, and learners’ experience compared to instructional videos. Video-based instruction was chosen as a consistent and easily replicable condition. By isolating visual demonstration from embodied instruction, the video format provides a stable baseline that minimizes variability in instructor behavior and classroom context, while still reflecting a practical training tool commonly used in vocational settings.

The study protocol was reviewed and approved by the Institutional Review Board (IRB) of our institution. Participation was voluntary; participants could withdraw at any time, and they received a \$50 Amazon gift card as compensation for their time. All data were anonymized during analysis to protect participant confidentiality.

\subsection{\textbf{Participants}}
We recruited 26 novice participants (13 male, 13 female), all adults aged 18-35 years old and from our institution. We excluded 1 male and 1 female in the data analysis process as their data were incomplete. None had prior welding training, reducing confounds from baseline skill.
\revised{\subsection{\textbf{Safety Procedures}}}
We worked closely with the welding training facility and our institutional review board (IRB) to establish safe conditions for all participant sessions. All participants completed a mandatory 30-minute safety training with the on-site welding instructor before engaging in any welding tasks. Each study session was also monitored by the welding instructor and a trained research staff member to provide oversight and respond to any issues during live welding.

\revised {Participants wore full welding PPE throughout the study, and the experimental setup was configured to minimize risk in the confined welding booth. We prepared two sets of controllers and headsets to avoid overheating from prolonged proximity to the torch.}

\revised{Prior to participant testing, we conducted validation experiments with welding instructors to confirm reliable system behavior, including assessments of pass-through latency, lighting variation, sensor accuracy, and tolerance to heat exposure. The Quest 3 pass-through used in our system offers low latency (40–60 ms)\citep{metaConsiderations}, allowing users to maintain near-real-time awareness of their environment—an essential requirement for live welding tasks. Across all sessions, no safety incidents occurred.}

\subsection{\textbf{Apparatus and Materials}}
\begin{figure*}[!htbp]
\centering
\includegraphics[width=\linewidth, alt={A participant wearing an AR welding helmet and protective gloves is actively welding on a steel workbench. A welding torch is held against the metal surface, with a small microphone positioned on the table nearby to capture audio. Several calibration tools and measurement instruments are arranged further along the workbench. Magnetic clamps and sample workpieces are fixed to the steel plates. A dedicated work light stands at the far end of the table, illuminating the welding area.}]{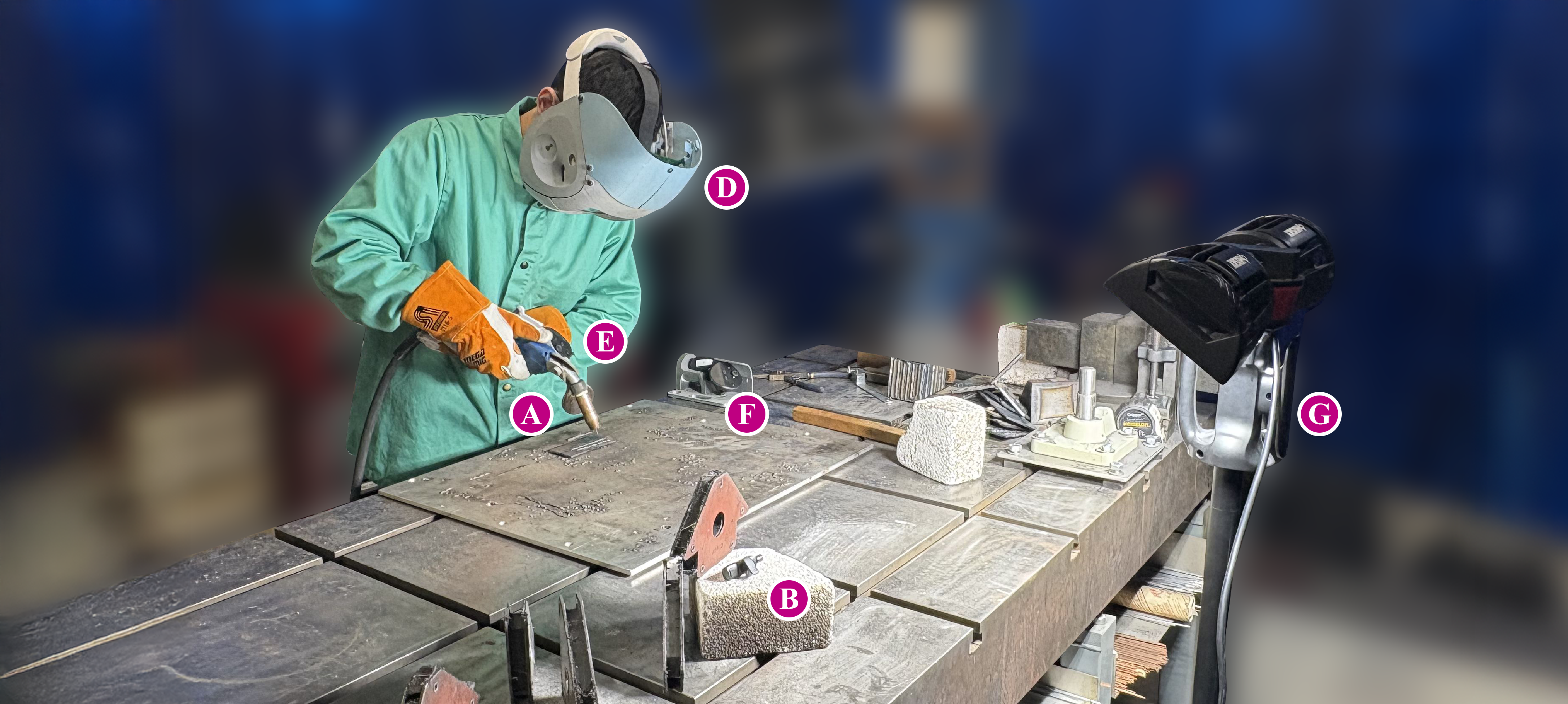}
\caption{Equipment shown: (A) Miller MDX-250 MIG weld torch; (B) Bluetooth lavalier microphone (microphone usb 1.0); (D) AR welding helmet with Meta Quest 3 integrated; (E) right controller (Quest Touch Pro) inside a bracket attached to the welding torch; (F) calibration station with left controller (Quest Touch Pro) inside; and (G) work light for illumination of the working area.}
\label{fig:welding_booth}
\end{figure*}

The study took place in a welding booth at our institution. Figure~\ref{fig:welding_booth} depicts the equipment used in the study. Participants wore the AR helmet (C, with auto-darkening screen) and welding Personal Protective Equipment (PPE) during the entire study. We used a Millermatic 212 Auto-Set, welding gun was MDX-250 (A), and the filler metal was 0.035" copper-coated mild steel. The welding coupons were 2" x 4", 11 gauge (rough $\frac{1}{8}$" or 3.04mm) mild steel, consistent with our institution’s training materials. 
A Bluetooth lavalier microphone (B) was placed 0.5 meters away from participants to capture audio during welding, a right-hand Quest Touch Pro controller inside a 3D printed bracket (E) was attached to the welding gun for motion data capture purposes, while the calibration station (F) was clamped to the edge of the workbench for setup. Additional lighting was provided by a standing work lamp (G) to ensure consistent visibility of the working area. \revised{This configuration matched the validation setup in Section 3.3 and maintained reliable tracking under the live arc}.

Within the AR system, we provided real-time visual feedback on four fundamental welding parameters: CTWD, travel angle, work angle, and travel speed. For comparison, we also provided pre-recorded instructional videos covering the same techniques.

\subsection{\textbf{Procedure/ Study Design}}

We conducted a within-subject user study to evaluate the efficacy of video instruction and AR guidance conditions. Participants were randomly assigned to one of two sequences: in the \textbf{Video-first sequence}, 12 participants first completed the video condition before the AR condition, whereas in the \textbf{AR-first sequence}, the remaining 12 participants completed the AR condition first, followed by the video condition.  

\begin{figure*}[!htbp]
\centering
\includegraphics[width=\linewidth, alt={Flowchart of the within-subjects welding study comparing AR and Video instruction. All participants (N=24) first complete safety training (30 min), study introduction and consent (10 min), and a workspace and tool overview. They then follow one of two counterbalanced sequences. In the AR-first sequence (N=12), participants complete AR training consisting of CTWD 4 lines, travel angle 4, work angle 4, speed 9, and combination 12, followed by an AR test of 6 lines, a 10-minute break, and then Video training with CTWD 4, travel angle 4, work angle 4, speed 4, and combination 12, followed by a Video test of 6 lines. In the Video-first sequence (N=12), participants complete the same Video training and test first, then after a 10-minute break complete the AR training and test. Both sequences end with a survey and interview lasting about 20 minutes.}]{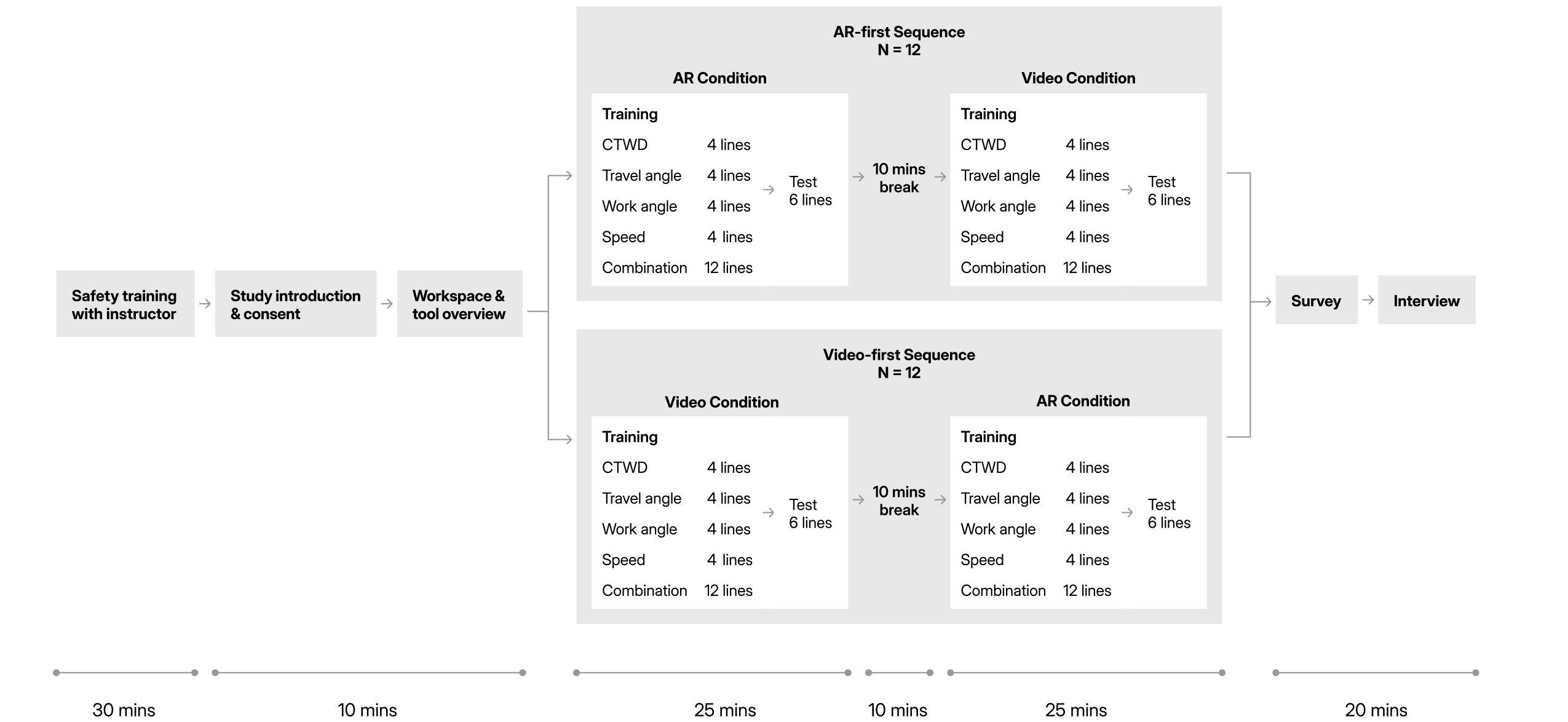}
\caption{The study employed a within-subjects design, where participants (N= 24) were placed in one of two groups. One group started with the AR training and test, followed by the Video training and test. Another group started with Video training, then a test, and then an AR training and test. }
\label{fig:procedure}
\end{figure*}

The study followed a structured sequence (see Figure~\ref{fig:procedure}) lasting approximately 90 minutes per participant.
\begin{enumerate}
    \item \textbf{Consent and Orientation (10 minutes)}: Participants were first informed of the overall study structure and signed a consent form, followed by an introduction to the welding booth and the equipment. They familiarized themselves with the AR helmet and the welding torch before starting the experiment session.
    
    \item \textbf{First experiment session (25 minutes)}: This was the session where participants finished their first condition. Participants completed four learning modules (CTWD, travel angle, work angle, and travel speed) followed by a practice combination module. Each learning module required participants to weld four lines, while the practice combination module required 12 lines. These repetitions were based on motor learning research that demonstrated the importance of repeated trials for early skill acquisition \citep{Johnson2023GeneralizationOP}.
    \begin{itemize}
        \item \textbf{Video Condition}: In each module, participants watched an instructional video to learn each technique and then practiced welding unassisted. After completing all modules, participants immediately performed an additional six lines unassisted as a test.
        \item \textbf{AR Condition}: In each module, participants practiced welding with real-time visual feedback on each technique and learned through hands-on experience. After completing all modules, participants immediately performed an additional six lines unassisted as a test.
    \end{itemize}
    
    \item \textbf{Break (10 minutes).} 
    
    \item \textbf{Second experiment session (25 minutes)}: Participants completed their second condition. Participants who completed the video Condition in the first experiment session then proceeded to the AR condition, and vice versa.    
    \item \textbf{Survey and Interview (20 minutes)}: Participants completed surveys on usability, workload, and perceived learning support, followed by a short semi-structured interview about both conditions.
\end{enumerate}

\subsection{\textbf{Data Collection}}
We collected three complementary types of data: welding performance measures, survey responses, and semi-structured interviews.

\subsubsection{\textbf{Welding Performance Data.}} 

\revised{The system recorded torch motion at 90 Hz using the Meta Quest Touch Pro controller. For every frame, we captured the 6-DOF pose and computed four welding parameters: CTWD (mm), Travel Angle (°), Work Angle (°), and Travel Speed (IPM). Each entry was labeled with participant ID, condition, module type, and timestamp. Participants completed 68 weld lines (34 per condition): 32 learning lines, 24 combination lines, and 12 final test lines. A small number of lines were discarded due to sensor drift during tracking or external acoustic noise.}

\revised{The AR scene was created in Unity 2022.03.15f1 and rendered on a Meta Quest 3 paired with Quest Pro controllers. The headset ran fully untethered, with the scene live-streamed through Quest Link. To maintain reliability under this wireless setup, the headset transmitted data via WebSocket to a local Windows 10 laptop (AMD Ryzen 7 5800H, RTX 3070). All logs were stored as structured JSON and mirrored to cloud storage, and physical welding coupons were retained for visual inspection.}

\subsubsection{\textbf{Survey Data.}} 

After both sessions, participants completed a survey derived from the Technology Acceptance Model (TAM) \citep{davis1989TAM} and the NASA Task Load Index (NASA-TLX) \citep{Hart1988DevelopmentON}. TAM was used to evaluate WeldAR as a training system in terms of perceived usefulness, ease of use, engagement, and behavioral intention. NASA-TLX was used to complement the objective performance metrics by assessing the systems’ mental and physical demands, perceived effort and workload, self-evaluation of performance, and frustration, reflecting the embodied nature of welding. 

The final survey instrument was a modified version of TAM and NASA-TLX, adapted for welding-specific contexts and is provided in Appendix ~\ref{app:survey}. A 7-point Likert scale was used for all questions. 

\subsubsection{\textbf{Interview Data.}}
To contextualize survey results, participants also completed a short semi-structured interview. The interview questions were grouped into five areas (see Appendix~\ref{app:questionnaire}):

\begin{itemize}
    \item General Experience: overall impressions and shifts in perception of welding.
    \item Experience During the Session: attention to real-time visual feedback, ease of the modules, challenges, moments of support or distraction, and equipment use.
    \item Impact on Performance: reflections on how each condition (AR and video) influenced task completion, memory of techniques, and confidence.
    \item Reflection and Evaluation: comparative perspective of usefulness and future adoption intentions of the system.
    \item Suggestions: recommendations for system improvements.
\end{itemize}

\subsection{\textbf{Data Analysis}}

\paragraph{\revised{Preprocessing and Metrics.}}
\revised{For each valid weld line, we calculated the mean absolute deviation of the trajectory frames from the expert-defined target ranges: CTWD $\in [6,15]$\,mm, Travel Angle $\in [-10,10]^{\circ}$, Work Angle $\in [75,90]^{\circ}$, and Travel Speed $\in [15,25]$\,IPM. To normalize performance across units, deviations were then converted to z-scores based on pooled statistics computed separately for the unassisted \textit{Test} and assisted \textit{Combination} sessions. We derived two participant-level metrics from these line-level z-scores:}
\begin{itemize}
    \item \revised{\textbf{Composite Performance:} The average z-score across all four parameters, where lower values indicate better adherence to targets.}
    \item \revised{\textbf{Stability:} The standard deviation of the composite z-scores across valid lines within a condition, representing consistency.}
\end{itemize}
\paragraph{\revised{Learning Trajectories.}}
\revised{We additionally tracked the composite z-scores across all sequential study segments to characterize the learning process. These data were aggregated to visualize performance trends and convergence patterns over time. This analysis was descriptive, intended to observe the nature of the learning curve.}

\paragraph{Survey Analysis.} We analyzed 24 participants’ responses using independent-samples t-tests to compare AR and video conditions across constructs of system acceptance (usefulness, ease of use, engagement, and behavioral intention) and workload (NASA-TLX subscales). Mean scores and standard deviations were reported for each measure, with significance levels (p-values) used to identify meaningful differences between the two conditions. This allowed us to capture both the statistical strength of differences and the practical implications of how participants experienced the training conditions.

\paragraph{Interview Coding and Analysis.} We conducted and transcribed a total of 452 minutes of interviews (7.5 hours) across 24 participants. The transcripts were imported into NVivo for analysis, where we applied a deductive coding framework \citep{fereday2006deductive} based on the aforementioned survey constructs. This approach enabled us to organize qualitative data in alignment with quantitative measures systematically and to capture participants’ nuanced reflections within each category.

\section{Findings}
\revised{We analyzed user study data regarding test performance, performance trajectory, and task load using a mixed-design Analysis of Variance (ANOVA) with Aligned Rank Transformation (ART), as data violated the normality assumption (Shapiro-Wilk test $p < .05$). The model examined the effects of the within-subjects factor \textsc{Condition} (2 levels: AR, Video) and the between-subjects factor \textsc{Group} (2 levels: AR-first, Video-first). We tested for main effects and interactions, conducting pairwise post-hoc comparisons with Bonferroni adjustment ($\alpha = .0083$) when significant effects were found. All statistical analyses were performed using R. }

\revised{In summary, we found that AR training led to significantly better retention and skill transfer than video training. Specifically, the AR-first group maintained high performance even after switching to video, whereas the Video-first group exhibited a marked performance drop when AR support was removed. Qualitatively, participants reported that AR improved \textbf{accessibility} and \textbf{confidence} by providing a safe, game-like environment with immediate error correction. However, results regarding \textbf{training support} and \textbf{ergonomics} revealed trade-offs: while the structured guidance was valued, participants occasionally struggled with visual information overload and physical fatigue due to the headset weight.}

\subsection{Quantitative Results}

\begin{figure*}[!htbp] \includegraphics[width=\linewidth, alt={A set of boxplots summarizing composite z-scored deviation for assisted combination sessions and unassisted test sessions across four sequence--condition groups. The x-axis lists the groups defined by training sequence and condition (AR-first/AR, AR-first/Video, Video-first/Video, and Video-first/AR), with separate boxes for the combination and test sessions within each group. The y-axis shows composite z-scored deviation, where lower values indicate better welding performance. Each boxplot depicts the distribution across participants, with central tendency and spread making it possible to compare performance patterns between sequences, conditions, and sessions.}]{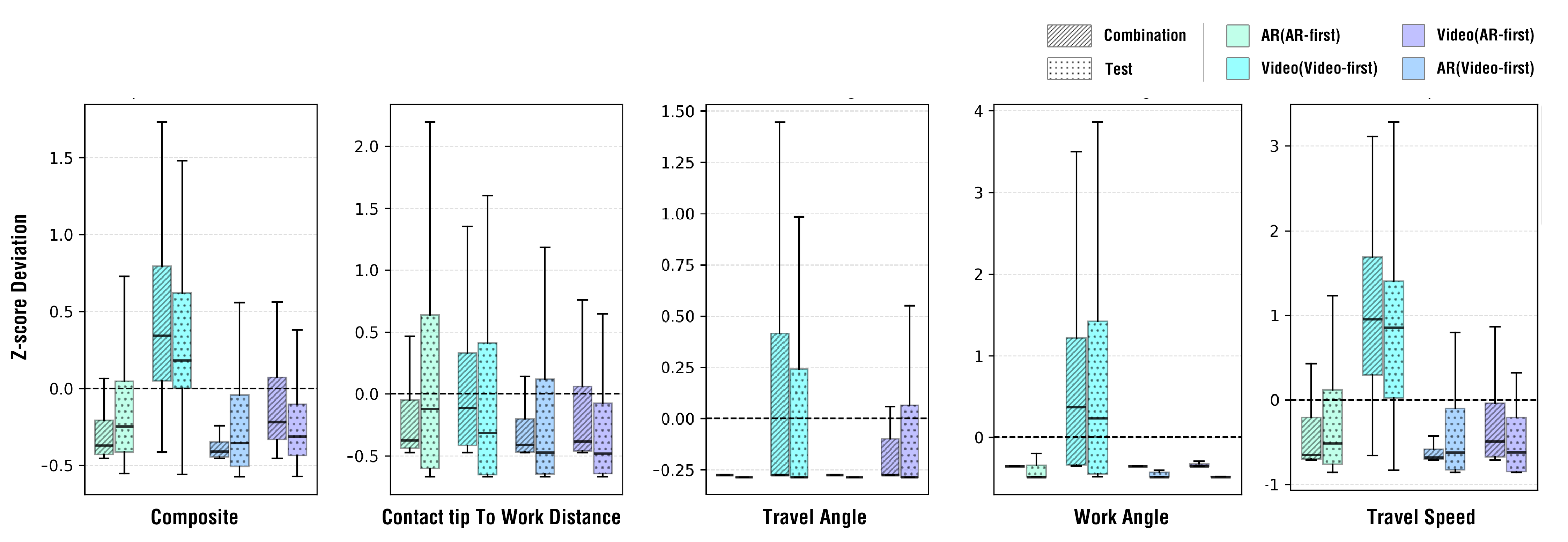} \caption{\revised{Summary of overall performance in assisted combination and unassisted test sessions across sequence and condition.}} \label{fig:chart_1_2} \end{figure*}

\revised{Performance results (Figure~\ref{fig:chart_1_2}, Figure~\ref{fig:chart_3}) demonstrate that AR training led to significantly better retention than video. Specifically, the AR-first group maintained high performance even after switching to video, whereas the Video-first group exhibited a marked performance drop when AR support was removed. Detailed datasets are provided in Appendix~\ref{app:performance_data} (Tables ~\ref{tab:perf_summary}, ~\ref{tab:zscored_deviation}, \ref{tab:participant_zscores_combined}}).

\subsubsection{\revised{Test session performance (Unassisted)}}

\revised{Analysis of composite deviation revealed a main effect of \textsc{condition} ($F_{1,22}=23.40,~p<.001,~\eta^{2}=.52$) and a \textsc{condition*group} interaction ($F_{1,22}=34.81,~p<.001,~\eta^{2}=.61$). Post-hoc comparisons regarding \textit{initial unassisted performance} revealed that participants who had just trained in AR performed significantly better in their first test than those who trained in video ($p = .032$). Regarding retention, the \textit{Video-first} group performed significantly worse when tested in video compared to AR ($p < .0001$), while the \textit{AR-first} group maintained their performance across conditions ($p = 1.00$). Furthermore, in the video test condition, the \textit{Video-first} group performed significantly worse than the \textit{AR-first} group ($p = .009$), indicating that prior AR training led to better skill transfer to unsupported environments than video training itself. Within-test variability did not differ across conditions or groups ($p \ge .50$).}

\revised{Skill-level analyses identified Work Angle and Travel Speed as the primary drivers of these results. Work Angle showed a significant main effect of \textsc{condition} ($F_{1,22}=9.80,~p=.005,~\eta^{2}=.31$) and a \textsc{condition*group} interaction ($F_{1,22}=12.11,~p=.002,~\eta^{2}=.36$). Speed deviation showed even stronger effects for \textsc{condition} ($F_{1,22}=17.33,~p<.001,~\eta^{2}=.44$), \textsc{group} ($F_{1,22}=5.21,~p=.032,~\eta^{2}=.19$), and interaction ($F_{1,22}=26.64,~p<.001,~\eta^{2}=.55$). Post-hoc comparisons clarified that for the initial test, the AR-trained group exhibited significantly better control than the Video-trained group for both Work Angle ($p = .007$) and Speed ($p = .031$). The interaction was further driven by the \textit{Video-first} group's significant degradation in these skills when tested in video (WA: $p < .001$; Speed: $p < .001$), whereas the \textit{AR-first} group remained stable (WA: $p = 1.00$; Speed: $p = 1.00$). In contrast, CTWD showed no significant effects (all $p \ge .18$), and Travel Angle showed only a marginal trend for \textsc{condition} ($F_{1,22}=3.50,~p=.075,~\eta^{2}=.14$). Sample test welds are shown in Figure~\ref{fig:combined_visuals}-1.}

\subsubsection{\revised{Combination session performance (Assisted)}}

\revised{Results indicate a main effect of \textsc{condition} ($F_{1,22}=51.58,~p<.001,~\eta^{2}=.70$) and a \textsc{condition*group} interaction ($F_{1,22}=34.80,~p<.001,~\eta^{2}=.61$). Post-hoc comparisons regarding \textit{initial exposure} revealed that participants starting with AR performed significantly better than those starting with video ($p < .001$). Following the modality switch, the \textit{Video-first} group exhibited a drastic performance drop when using video compared to AR ($p < .0001$), whereas the \textit{AR-first} group showed a much smaller decline ($p = .027$). Within-session variability showed a similar interaction ($F_{1,22}=11.21,~p=.002,~\eta^{2}=.34$): variability was significantly higher for the \textit{Video-first} group during their initial video exposure compared to the \textit{AR-first} group's initial AR exposure ($p = .13$, ns), but the Video-first group became significantly more unstable when using video compared to AR ($p < .0001$).}

\revised{Skill-level analyses identified Travel Speed and Work Angle as the primary drivers of these effects. Travel Speed exhibited the strongest results, with significant effects for \textsc{condition} ($F_{1,22}=76.63,~p<.001,~\eta^{2}=.78$), \textsc{group} ($F_{1,22}=9.89,~p=.005,~\eta^{2}=.31$), and interaction ($F_{1,22}=45.87,~p<.001,~\eta^{2}=.68$). Work Angle showed similar significance for \textsc{condition} ($F_{1,22}=20.24,~p<.001,~\eta^{2}=.48$), \textsc{group} ($F_{1,22}=7.23,~p=.013,~\eta^{2}=.25$), and interaction ($F_{1,22}=9.40,~p=.006,~\eta^{2}=.30$). For initial exposure, the AR-first group showed significantly better control than the Video-first group in both Speed ($p < .001$) and Work Angle ($p < .001$). The interaction effect was confirmed by the \textit{Video-first} group's significant difficulty in maintaining these parameters when using video (Speed: $p < .001$; WA: $p < .001$), while the \textit{AR-first} group remained stable (Speed: $p = .50$; WA: $p = .20$). CTWD presented a significant interaction ($F_{1,22}=13.78,~p=.001,~\eta^{2}=.39$) but no difference in initial exposure ($p=.19$). Travel Angle showed only a main effect of \textsc{condition} ($F_{1,22}=6.94,~p=.015,~\eta^{2}=.24$) with no significant interaction ($p=.125$).}

\subsubsection{\revised{Learning trends}}

We analyzed the learning rates and the immediate performance impact of switching training modalities. Regarding the \textit{initial training phase}, an analysis of linear regression slopes across the eight training segments confirmed that the \textit{AR-first} group exhibited a significantly steeper learning curve ($M_{slope} = -0.201$) compared to the \textit{Video-first} group ($M_{slope} = -0.075$; $p = .005$), showing participants who were trained under AR guidance acquired correct skills more rapidly than those with video instructions.

\begin{figure*}[!htbp]
\centering
\includegraphics[width=\linewidth, alt={A line chart showing mean z-scored deviation across training segments for two learning sequences: AR-first and Video-first. Each sequence contains eight segments in the first condition (CTWD, Travel Angle, Work Angle, Travel Speed, Combination 1--3, Sample test welds are shown in Figure~\ref{fig:test_result}.Test) followed by eight segments in the second condition, separated by a dashed vertical divider at the midpoint. The AR-first sequence is plotted in blue, and the Video-first sequence is plotted in red. The x-axis lists the ordered segments from both conditions, and the y-axis represents the mean composite z-scored deviation.}]{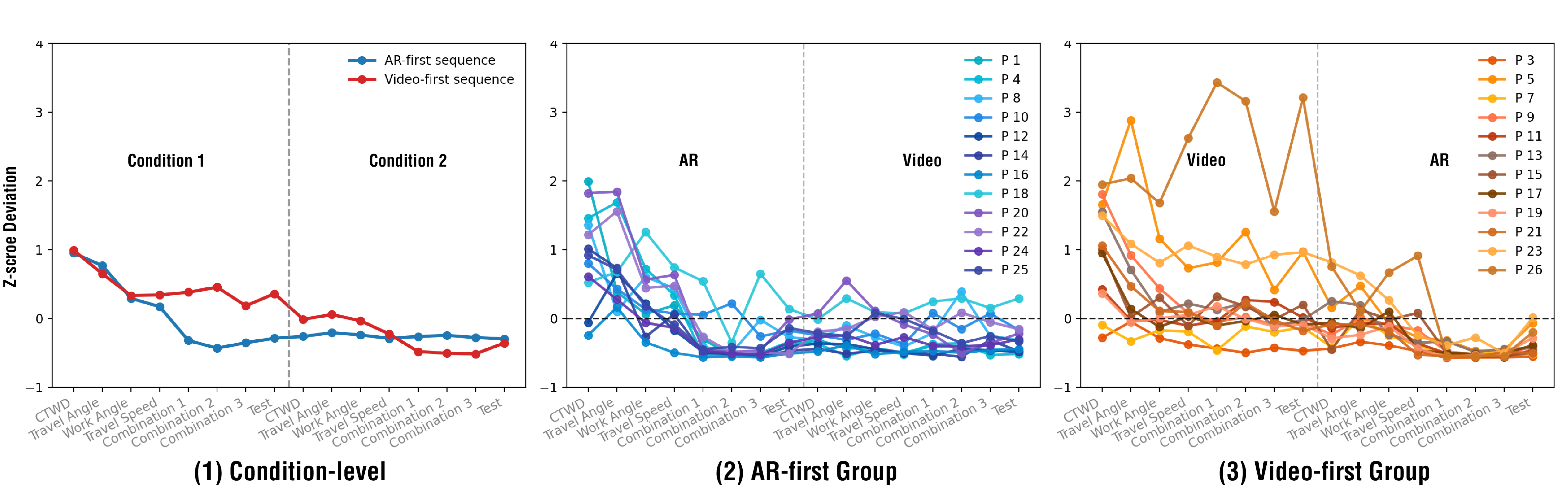}
\caption{\revised{Group--condition mean performance trajectories across the two training sequences. (1) Condition-level performance (2) AR-first group performance (3) Video-first group performance.}}
\label{fig:chart_3}
\end{figure*}

\begin{figure*}[!htbp]
\centering
\includegraphics[width=0.95\linewidth, alt={A composite figure divided into two panels. Panel (1) 'Test Session' on the left shows close-up photographs of weld plates from the final test. The 'Video-first' samples (top) display irregular, wavy beads, whereas the 'AR-first' samples (bottom) show consistent, straight beads. Panel (2) 'All Sessions' on the right presents vertical columns of weld plates corresponding to the full curriculum (CTWD, Travel Angle, Work Angle, Travel Speed, Combinations, and Final Test). The 'AR-first' columns show consistent quality starting with AR and maintaining it in Video. The 'Video-first' columns show inconsistent welds during the initial Video phase, which improve significantly after switching to AR.}]{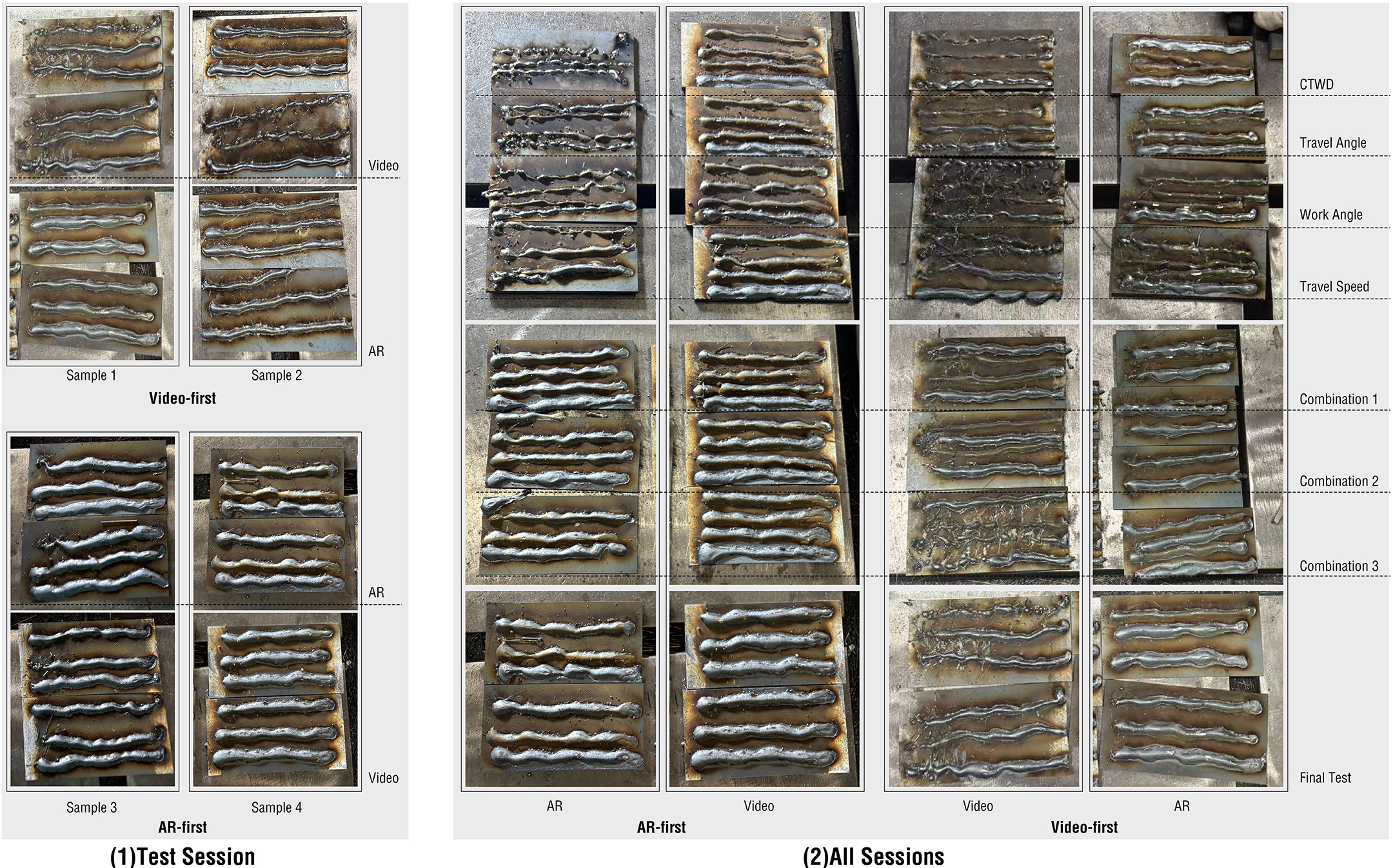}
\caption{\revised{Weld artifacts. (1) illustrates specific outcomes during the unassisted Test Session, highlighting the retention difference between groups. (2) displays the learning trajectory across all sessions, showing the progression of practice.}}
\label{fig:combined_visuals}
\end{figure*}

To further analyze performance changes upon switching conditions, we calculated the immediate change in deviation $\Delta$ between the end of the first condition (3 combination sessions) and the start of the second (distance and angle sessions). An independent samples t-test confirmed notable difference in the drop between the two groups ($p = .022$). Specifically, the \textit{Video-first} group exhibited a substantial performance improvement (a drop in deviation with mean $\Delta = -0.329$) when switching to AR guidance, whereas the \textit{AR-first} group showed a slight performance degradation (mean $\Delta = +0.121$) when switching to video. These results indicate that switching from Video to AR provides an immediate benefit distinct from the growth of deviation observed when switching from AR to Video. The condition-level and individual performance trajectories illustrating these trends are visualized in Figure~\ref{fig:chart_3}, with corresponding sample welds shown in Figure~\ref{fig:combined_visuals}-2.

\subsubsection{Systems Acceptance.} 
\begin{figure*}[!htbp]
\centering
\includegraphics[width=\linewidth, alt={A horizontal stacked bar chart showing the Technology Acceptance Model responses comparing AR (Augmented Reality) and Video training methods across nine dimensions. Each dimension shows two bars representing AR and Video conditions, with responses distributed across a 7-point scale from 1 (Strongly Disagree) to 7 (Strongly Agree).}]{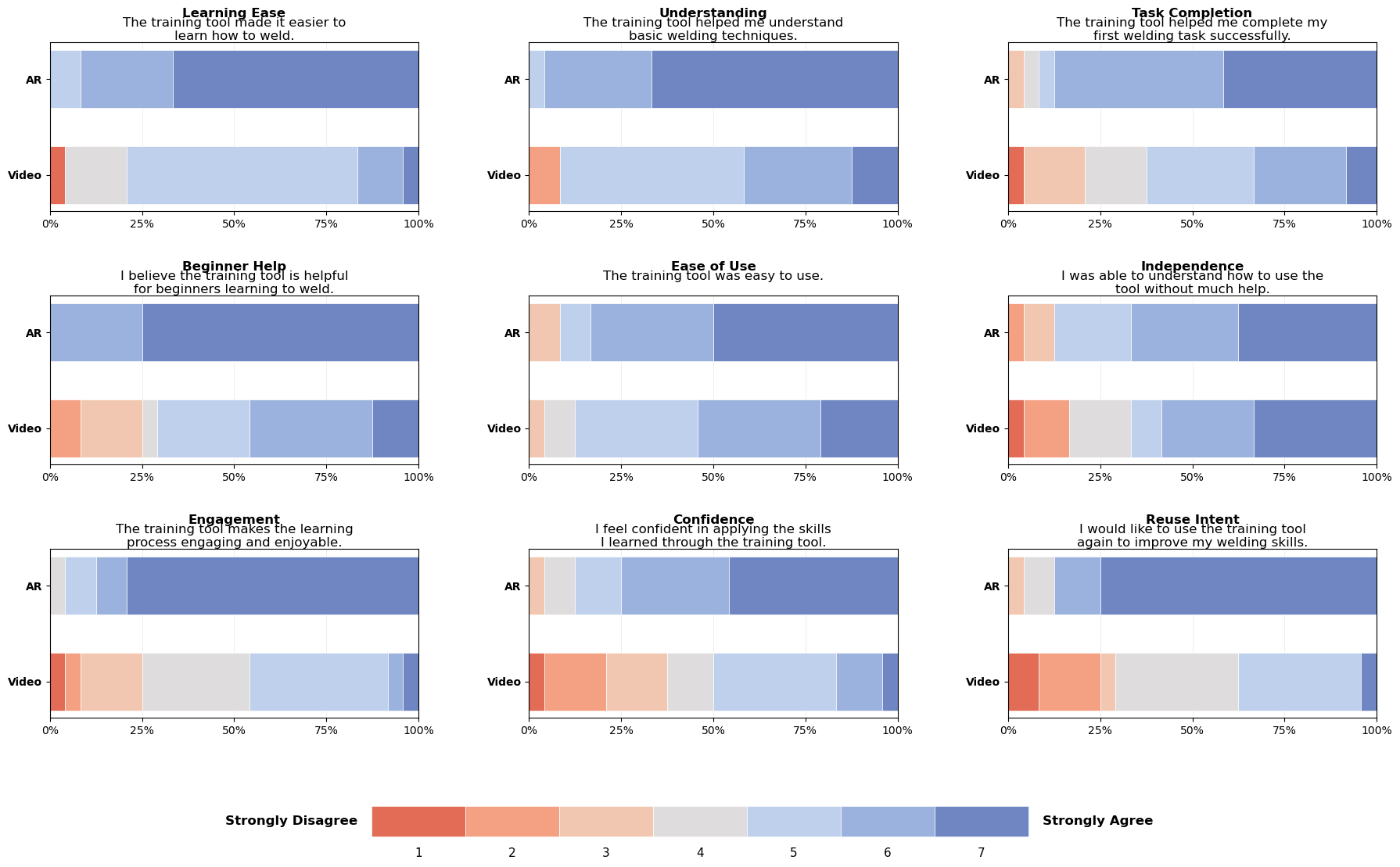}
\caption{Distribution of the Technology Acceptance Model responses, with each bar showing the proportion of participants (0-100\%) who selected each rating.}
\label{fig:tam}
\end{figure*}

The TAM was employed to assess perceived usefulness, ease of use, engagement, and behavioral intention between AR and Video training (see Figure~\ref{fig:tam}).

\paragraph{Perceived Usefulness.} In our study, Perceived Usefulness captures whether participants felt the tools supported their ability to learn welding techniques and accomplish training tasks. AR training was rated as more useful than Video across multiple items. Participants felt that AR made welding easier to learn (M = 6.54, SD = 0.66) compared to Video (M = 4.87, SD = 1.07), a significant difference (p < .001). Similarly, AR was rated higher in helping participants understand basic welding techniques (M = 6.68 vs. 5.29, p = .0006) and supporting successful task completion (M = 6.13 vs. 4.75, p = .002). 

Beyond immediate learning tasks, AR was rated as more helpful for beginners overall (M = 6.75 vs. 4.95, p < .0001), and participants expressed greater confidence in applying welding skills after AR practice (M = 6.04 vs. 4.00, p < .001). These results suggest that AR provided a stronger sense of mastery and practical applicability.

\paragraph{Perceived Ease of Use.} Ease of use scores did not differ significantly between AR and Video. Participants rated AR as slightly easier to use (M = 6.13, SD = 1.15) than Video (M = 5.653, SD = 1.06), but the difference was not significant (p = .090). Similarly, both groups felt they could use the training tool without much help (AR: M = 5.67 vs. Video: M = 5.38, p = .307). These findings indicate that AR, despite requiring a more unfamiliar format, did not present additional usability barriers compared to the familiar video format. 

\paragraph{Perceived Engagement.} Ratings for engagement were markedly higher for AR (M = 6.62, SD = 0.82) than for Video (M = 4.21, SD = 1.28), a highly significant difference (p < .001).

\paragraph{Behavioral Intention.} Participants reported a higher desire to reuse AR to improve their skills in the future (M = 6.45, SD = 1.14) compared to Video (M = 3.83, SD = 1.49, p < .001). Similarly, AR users were more likely to recommend the tool to others (M = 6.75, SD = 0.68) than Video users (M = 4.62, SD = 1.58, p < .001). 

\subsubsection{Task Load}
\begin{figure*}[!htbp]
\centering
\includegraphics[width=\linewidth, alt={A horizontal stacked bar chart showing NASA Task Load Index responses comparing AR (Augmented Reality) and Video training tools across six dimensions. Each dimension shows two bars representing AR and Video conditions, with responses distributed across a 7-point scale from 1 (Very High/Rushed/Failure) to 7 (Very Low/Relaxed/Excellent).}]{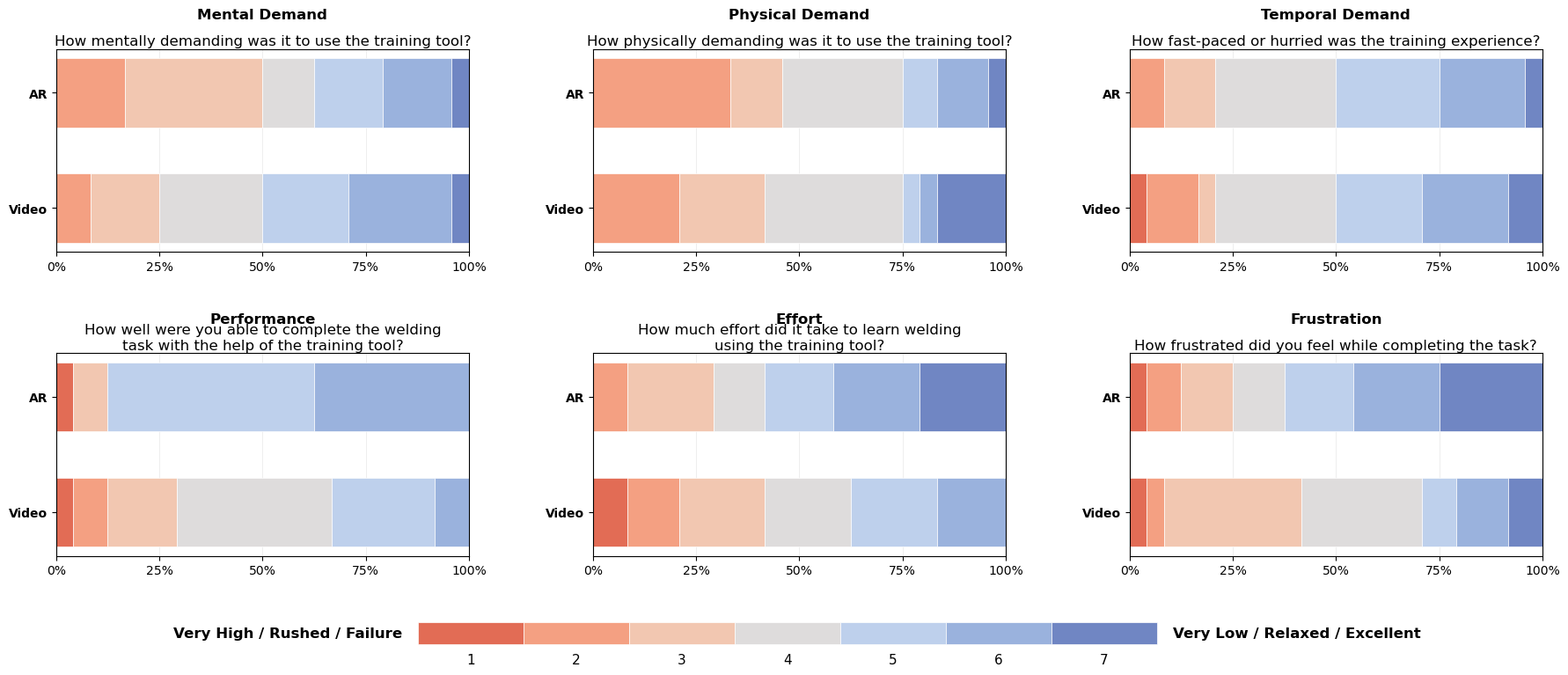}
\caption{Distribution of NASA Task Load Index responses, with each bar showing the proportion of participants (0-100\%) who selected each rating.}
\label{fig:nasa}
\end{figure*}

The NASA TLX ratings highlight differences in workload perceptions between AR and Video training (see Figure~\ref{fig:nasa}). Independent-sample t-tests were conducted to compare conditions across six subscales. A main effect of training condition was observed for performance, with AR (M = 5.00, SD = 1.22) rated higher than Video (M = 3.92, SD = 1.28; p = .0012). This suggests that participants perceived themselves as more successful in completing welding tasks with AR support. Significant effects were also found for questions on effort and frustration. AR scored higher on effort (M = 4.83, SD = 1.69) than Video (M = 3.71, SD = 1.57; p = .0008), suggesting that participants experienced welding with AR as requiring less exertion and being easier to manage. Similarly, AR participants reported higher scores on frustration (M = 4.88, SD = 1.68) compared to Video (M = 3.83, SD = 1.55; p = .024), indicating that they felt less frustrated when using AR. No significant differences were found for mental demand (AR: M = 3.95 SD= 1.51; Video: M = 4.5, SD = 1.38; p = .100), physical demand (AR: M = 3.66, SD = 1.55; Video: M = 3.66, SD = 1.55; p = .31), or temporal demand (AR: M = 4.5, SD = 1.31; Video: M = 4.45, SD = 1.61; p = .913). Although the average scores show slight differences across these subscales, the results were not statistically significant. This suggests that AR was not perceived as clearly less mentally, physically, or temporally demanding than Video.

\subsection{Interview results}
The interview findings provide additional context on the performance of AR relative to video. Three themes emerged: perceived accessibility, support for the learning process (particularly in building confidence), and overall training experience.

\subsubsection{Perceived Accessibility. }
Participants often described welding as intimidating before the study, associating it with heat, sparks, and heavy equipment. One participant thought welding was \textit{“scary, dangerous, and only for professionals”} (P17).  In AR sessions, ten participants reported that the structured guidance supported them in attempting the task: \textit{“It guided me precisely on what I should be doing, which made it feel less overwhelming and helped me stay calmer”} (P12).  Another also reported, \textit{“after practicing with AR, I can see it takes some skill, but I think it's achievable”} (P02).  In contrast, twelve participants said that videos were less helpful. One explained that after watching the lesson video, their weld lines became inconsistent, and it was \textit{“a little confusing figuring out the distance that you’re supposed to achieve" }(P03). Another commented that the lesson video \textit{“wasn’t helpful because even though I watched it and it was a completely different experience”} (P05). 

Thirteen participants described the AR interface as intuitive once they became familiar with it: \textit{ “Initially, I couldn’t both read and then understand what it was trying to give … but once I got it, the green and red feedback was helpful”} (P14).  Four participants characterized the AR interface game-like, describing the interactive feedback as similar to following rules in a game: \textit{“It’s like a game, more playful AR setting”} (P09).  However, one participant noted that the game-like qualities could create distance from welding as a physical task that requires carefulness: \textit{“it feels very much like a game. You’re not 100\% aware that you’re actually doing a dangerous and physically involved task. It just feels more like a game”} (P21). 

\subsubsection{Confidence in Performance}
Participants frequently described AR’s real-time guidance as improving their confidence by helping them identify and correct mistakes immediately across all four skill parameters. One participant said \textit{“…when I did AR, I feel like they have the real-time feedback, like the distance or the speed or the angle, so that I can see the feedback right away. So it was beneficial to understand how much I have to do for the right welding”} (P5).  Another commented, \textit{“I think AR had a lot of useful feedback, and as a beginner, I needed that to understand what it was all about”} (P20). When describing video, seventeen participants noted that it provided a model of the “correct” technique but did not help them evaluate their own performance. One participant said, \textit{“I actually don’t know [whether my techniques] right or wrong until the end”} (P17).  Another summarized, \textit{“Video just gave me the right example, whereas the AR showed me the way to do it”} (P22).

\subsubsection{Perceived Training Support}

\paragraph{Real-time UI Feedback.} Participants described AR's real-time feedback as both helpful and effortful to manage. The two color-coded system (green and red) was noted as a clear guide, but some participants said that when four visual indicators appeared at once, it was hard to follow them all: \textit{“I’m just trying to keep everything green, that’s all I can pay attention to”} (P24). Nine participants said the red signals created anxiety. One described, \textit{“…whenever I felt something was turning red, I just stopped, [myself] panicked. I don’t know if that had to be done or not. Initially, that happened a lot”} (P23). Others said that the sudden full-screen red speed indicator felt distracting, comparing it to a \textit{“hazard signal”} (P15).

Twelve participants also commented on the placement of UI (speed indicator) feedback, wanting to have the speed indicator appear within their field of view. As the speed indicator was placed at the lower center field of view it may require head movement to see it clearly; participants said this made it harder for them to follow during welding: \textit{“[Speed indicator] was there while I was in the middle of welding, but I couldn’t really see it. I had to look and mentally correlate it, and that was hard”} (P33).


\paragraph{Hardest and Easiest Techniques.} In the interviews, participants identified differences in difficulty across the four welding techniques. Eighteen participants described both angles as easier to manage. One explained, \textit{“I’d say both of the angles were pretty easy, although I think that the travel angle… was probably the easiest”} (P02). Another said, \textit{“For the two angles, I just need to hold my hand still”} (P18). A third participant commented,\textit{ “I thought the angle was the easier part, at least based on the results that I was getting”} (P12).

Twenty participants described travel speed and CTWD as the hardest techniques. One stated, \textit{“I’d say speed, purely because… I’m not used to the motions [of welding]”} (P03). Another reported, \textit{“It was actually way harder to maintain a constant speed than I expected”} (P07). A participant explained, \textit{“The distance… I couldn’t really see the little plate. So then… it was hard to know how far away”} (P06). One participant summarized, \textit{“The speed is the hardest… because angle you could fix it, but for speed… combining both [speed and CTWD] was hard”} (P22). In general, both angles were perceived to be easiest, as they involve holding the welding torch still, while speed and CTWD were perceived as are harder because they require continuous adjustment during welding.

\subsubsection{Skill Development Support}
\paragraph{Beyond visual real-time feedback.} Participants described how AR training gradually shifted their attention from the on-screen indicators to the underlying sensory cues of welding. Repeated practice led them to notice differences in welding sounds, vibrations of the welding torch, and sometimes heat from the welding coupon. Twenty participants mentioned sound as a physical cue. One said, \textit{“…for the AR, I didn’t take a lot of help from the visual cue after some time… but it was more of the sound of it. If I’m at the right distance, I’m getting continuous [sound]”} (P22). Nine participants described attention to the feel of vibration and stability in the welding torch, using it as a marker of correctness: \textit{“…because when you do it, you kind of get the sense of the sparks that cause the vibration of your hand, so you might kind of know if it’s going right or wrong”} (P19). Three participants noticed heat as a diagnostic cue: \textit{“It became more about sensing, noticing the sound of the arc and even how hot it felt, to judge whether my speed was right”} (P22).  However, six participants said that once they began to pay attention to welding sounds, vibration, and weld pool color, the AR-based feedback and overlays became distracting. One explained, \textit{“In the beginning, the lines and colors were really helpful to [understand] the angles and distance. But once I started paying more attention to how the plate reacted, the overlays felt more like they were always there so I can not see clearly what my plate looks like–especially because of the distance feedback”} (P18).

This contrasted with video-based training, where participants relied on mimicking expert movements and checking outcomes. One said, \textit{“…afterwards, when I looked at my line, it wasn’t solid, so I knew it wasn’t good”} (P15). Four participants acknowledged a benefit from watching demonstrations: \textit{“there is something [valuable] for watching an expert do it, especially with respect to like how you hold your hands. In AR, I was receiving feedback that my angle was incorrect, but not on how to achieve the correct angle”} (P10).

\paragraph{Muscle Memory.} Twelve Participants described muscle memory as a link between guided training and independent tasks. Rather than strictly cognitive recall, they perceived the repetition as aiding physical retention. One participant said, \textit{“When I succeed, I can remember the right position and angles, and it shows my muscle memory. And when I failed, I was also trying to figure out… what’s the point that led to the failure? I become more and more aware of how I should pause it, what’s the right angle, and how to watch the trajectory”} (P1). Fifteen participants contrasted this with video-based training, which they said provided fewer scaffolds for developing consistent movement patterns. 

\subsubsection{Task Load and Ergonomics}

Participants described how the physical hardware of the AR system, including the helmet and welding torch attachment, shaped their welding experience. The AR helmet was described by 18 participants as heavy and front-weighted, contributing to muscle strain during practice. \textit{“When I feel tired, my neck hurts and I just [get] distracted… the weight is on the front of the head, so it makes me feel more tired”} (P6). Six participants compared the weight to a conventional welding helmet and noted that the imbalance felt uncomfortable. Six participants also mentioned that the welding torch attachment affected posture and was slightly obstructive to natural welding motions. \textit{“I found the controller to be, like, slightly obstructive of the actual tool that you’re using. You’re sort of forced to either do two hands like this (showing the motion)”} (P11). Another explained, \textit{“I don’t think it was just a heavy gun…  it’s just kind of unnatural”} (P12).

Visual challenges during welding were described as constraints. Sixteen participants reported difficulty seeing the welding coupon or aiming accurately due to low visibility while wearing the helmet. One explained that the welding coupon was \textit{“not visible, and then the dot doesn’t help, because it kind of like covers over the tip”} (P11). Similarly, one reported that when \textit{“I cannot see the edge and cannot feel that. So sometimes I go off the coupon, I don’t realize it”} (P13). \revised{These challenges reflect the difficulty that beginners often face when first learning to visually navigate through darkened welding helmets, rather than an AR-specific limitation.}

\begin{table*}[ht]
\caption{Summary of research questions (RQ) and key findings from the AR welding study.}
\label{tab:rq_findings}
\renewcommand{\arraystretch}{1.2}
\begin{tabular}{p{0.25\linewidth} p{0.70\linewidth}}
\toprule
\textbf{RQ} & \textbf{Findings} \\
\midrule

RQ1: Effect of AR on novice performance (CTWD, angles, speed) &
\begin{itemize}
  \item AR-first outperformed Video-first on travel/work angles and travel speed; CTWD similar.
  \item Overall AR performance was more accurate and stable than video.
  \item Sequence effect for Video to AR improved substantially; AR to Video stayed stable.
  \item Robustness check confirmed trends after removing outliers.
  \item Interviews revealed AR feedback (green/red cues) supported immediate correction of distance, angles, and speed; however, simultaneous cues sometimes led to cognitive overload.
\end{itemize}\\

\addlinespace

RQ2: AR support for transfer to independent, no-feedback welding &
\begin{itemize}
  \item AR-first group maintained stable performance when switching to video, indicating successful skill transfer.
  \item Video group in Video-first still improved, but at slower rate and without convergence to AR-level performance. 
  \item Participants reported shifting attention from AR overlays to embodied cues, including welding sounds, torch vibration, and heat.
  \item AR practice built muscle memory, whereas video-based instruction primarily provided demonstrations with limited support for self-evaluation.
\end{itemize}\\

\addlinespace

RQ3: Ergonomics, usability, experiential factors &
\begin{itemize}
  \item Perceived usefulness shows AR rated significantly more helpful than video.
  \item Ease of use shows no significant difference, AR was perceived the same despite novelty.
  \item Engagement shows AR as substantially more engaging than video.
  \item Behavioral intention shows participants wanted to reuse/recommend AR more than Video.
  \item Task load factor: AR perceived as more successful, less effortful, less frustrating; no differences in mental/physical/temporal demand.
  \item Accessibility: AR reduced intimidation, made welding feel achievable; Video less helpful.
  \item Training support with structured modules is perceived to be helpful; though UI placement and simultaneous cues occasionally caused overload.
  \item Ergonomics factor: AR helmet heavy/front-loaded, causing neck strain; the torch-mounted controller sometimes constrained natural movement; visibility issues limit accuracy.
\end{itemize} \\

\bottomrule
\end{tabular}
\end{table*}

\section{Discussion}
\subsection{How AR Shapes Early Learning of Live Welding}

\paragraph{Assistance in MIG weld learning.} \revised{In the first session, AR yielded lower composite deviation scores and faster improvement than video, particularly in speed and work angle.} Participants exposed to AR showed earlier gains in performance and a lower rebound, while those starting with video showed less improvement and more rebound after the test module. Real-time feedback and post-module summaries provided actionable guidance, helping to stabilize performance and build up confidence. \revised{Notably, the AR-first group showed zero performance decay ($p=1.00$) when guidance was removed, indicating that AR facilitated true skill internalization rather than serving as a temporary crutch.} Interviews further suggest that AR cues helped learners build muscle memory. They not only adjusted in real time but also retained the 'feel' of correct posture and movement. Over practice, participants described gradually shifting attention from visual to embodied cues such as sound, heat, and vibration, indicating that AR seeded motor routines that persisted into subsequent practice and testing. Beyond performance gains, AR reframes assessment from post-hoc coupon inspection to formative, in-booth real-time feedback.

\paragraph{Consistent benefits of AR guidance.} AR consistently supported improvement regardless of order. When AR followed video, sharp improvements were observed; when AR preceded video, performance still improved gradually. This robustness against sequence order suggests that AR provides a reliable beneficial effect. Interviews also showed that in both sequences, participants felt more confident in their performance when practicing with AR, describing the feedback as reassuring and supportive of their progress.

\paragraph{Different effects across parameters.} From the quantitative results, the strongest gains under AR were seen in Travel Speed, followed by Travel and Work Angles, whereas CTWD improvements were limited. Interviews, however, noted that angles were easier because they involved holding the torch relatively still, while speed and CTWD required continuous adjustment. Participants reported that keeping speed consistent was particularly challenging without real-time cues, which helps explain why AR support yielded the largest advantage here. CTWD, by contrast, remained harder to improve. This was largely linked to the visual limitations of the headset, as interviews pointed to distortion, low resolution, and limited depth perception, which together made it difficult sometimes to see the coupon clearly or judge distance accurately. Although AR generally helps better with dynamic skills than with static ones, these issues suggest that display fidelity constrains performance gains for certain parameters.

\paragraph{Comprehensive feedback.} The combination module yielded the largest performance gains, where all four cues were presented together. Comparing the overall performance trajectories between the two conditions, the deviation in the AR condition dropped more rapidly after the combination module than after other modules. Our findings suggested, however, participants described the full set of indicators as overwhelming, especially the speed signal, which sometimes caused them to pause or feel overloaded. This tension suggests a tradeoff: while comprehensive feedback increased momentary workload, it also pushed learners to coordinate multiple parameters simultaneously, accelerating \textit{learning consolidation}, the process of stabilizing new skills into lasting performance. This reveals tensions between quantitative evidence and user feedback: while metrics show rapid consolidation, subjective reports emphasize overload. Taken together, the results suggest that experiences perceived as demanding during training may ultimately contribute to more resilient skill acquisition.

Potential visual occlusion from overlays represents a tested design trade-off. Prior pilot studies (workshops conducted at our partner site) and feedback showed that overlays placed farther from the torch increased cognitive load, while torch-mounted overlays occasionally caused mild occlusion. We therefore adopted a torch-mounted design with adjusted opacity to balance visibility and line of sight, with occlusion concerns primarily emerging during the user study rather than earlier piloting.

\subsection{Designing AR for Fostering Embodied Learning}
Reflecting on our findings, we articulate what it means to design for embodied learning in this context. Embodied learning comprises three dimensions: sensorimotor skill, perceptual knowledge, and situational awareness \citep{polanyi_tacit_2009,chatain_three_2023}. Across these dimensions, our findings suggest that the design of AR training must not only scaffold technical correctness but also support learners in cultivating embodied control, perceptual attunement, and safe, situated practice.

\subsubsection{From Judgments to Embodied Control.} Skill acquisition in physical domains is grounded in sensorimotor integration. Classic models of motor learning emphasize that learners internalize control by coupling perception and action through iterative cycles of practice \citep{schmidt2011motor}, gradually constructing forward models and refining proprioceptive awareness \citep{kawato1999internal}. Welding exemplifies this process: novices must coordinate distance, angles, and travel speed within narrow tolerances, where small deviations compromise both weld quality and safety.

For this reason, effective training should help learners build confidence in their ability to maintain control within these tolerances instead of prioritizing binary correctness. Our findings illustrate this distinction. Some participants described heightened stress when confronted with persistent red/green (right/wrong) overlays, suggesting that binary cues displaced attention from embodied sensing and undermined confidence in controlling their movements. While easily interpretable, such feedback can oversimplify a complex practice into right versus wrong rather than scaffolding learners’ ability to continuously regulate technique.


\subsubsection{Moving Beyond Sensorimotor Towards Perceptual Knowledge}
The second dimension of embodied learning emphasizes perceptual skill, where learners move beyond stabilizing motor patterns to “see” and “feel” the practice differently \citep{polanyi_tacit_2009}. In welding, this involves recognizing tacit cues such as the sound of the arc, the vibration of the torch, and the bead's appearance. Because real welding often limits visibility through glare, sparks, and protective equipment, welders must rely heavily on these perceptual channels to guide their actions. This aligns with vocational welding instruction, where instructors frequently highlight the importance of embodied cues. 

Participants in our study described this transition: they emphasized the need for clear feedback at the outset to verify the correctness of their technique. Over time, however, they described this constant AR feedback as distracting, since it pulled their attention away from the welding plate, bead, and process. Several noted that during the exam, when the UI was removed, they relied instead on physical cues such as sound, vibration, and the stability of the torch to guide their performance. Notably, this shift occurred without explicit instruction, suggesting that learners naturally began to rely on embodied signals once feedback was reduced. Participants expressed a preference for feedback that gradually fades, allowing them to shift focus toward these signals for effective transfer. 

The critical question is how to design the transition from sensorimotor control to perceptual attunement. Prior research on motor skill learning shows that faded feedback, where knowledge of results (KR) or augmented feedback is provided less frequently, improves long-term retention and transfer \citep{aoyagi2019fading}. This approach may be valuable in welding as well, but welding places unique demands on the process because competence depends not only on technical correctness but also on sensitivity to perceptual cues that guide safe and effective performance.

\subsubsection{Balancing Accessibility and Embodied Practice}
Embodied learning extends beyond skill execution to include situatedness and social context \citep{Hadjimichael2024TacitEmbodiment}. Prior work emphasizes that tacit knowledge emerges not only from repeated movements but also from engagement with the environment and participation in shared practice \citep{chatain_three_2023}. This includes orienting the body, interpreting sensory cues, and situating action within broader material and social conditions rather than technique alone. Welding illustrates this clearly. Competence extends beyond mastering parameters to include orienting within the booth, maintaining ergonomic posture, and handling materials. Crucially, in welding, these embodied dimensions matter not only for learning but also for ensuring safety \citep{chen_augmenting_2024}.

Our findings highlight this importance. Our system focused on four welding parameters, presented through overlays, real-time scoring, and interactive feedback that made training more accessible and approachable. While participants noted that this lowered barriers to entry and encouraged exploration, it also risked shifting attention away from risk awareness, precision, and embodied cues essential for developing tacit knowledge. When learners were oriented toward overlays rather than the broader environment, aspects such as posture, material handling, and spatial awareness were less salient.

The implication for AR training in high-risk vocational domains is that lowering barriers to entry should not come at the cost of embodied rigor. This tension aligns with broader HCI research, which shows that embodied interaction alone does not guarantee embodied cognition to sustain meaningful learning \citep{chatain_three_2023}. The challenge, however, lies in sequencing this transition: how to move from accessible entry points toward embodied rigor that emphasizes attentiveness, orientation, and safety.

\paragraph{Implications for Vocational Training (Welding)}
AR gives immediate feedback that helps beginners understand what they are doing right and what needs improvement, thereby building confidence. This early confidence can help reduce welding's intimidation factor—often the biggest hurdle for new students. With increased confidence, learners are likely more willing to practice, experiment, and gradually develop the physical control that skilled welding requires. For instructors, confident students can reduce teaching pressure: when instructor time is limited, students need less constant guidance, and since welding is often seen as difficult, confident beginners are more likely to persist in the program. 

AR training helps beginners notice the physical aspects of welding. As they practice, they start to see how small changes in hand movement affect the quality of their welds. Because skilled welding depends on reading these physical cues, and because such knowledge is difficult to teach through verbal instruction alone and often requires extensive hands-on practice, AR may help bridge this gap by allowing beginners to experience what good welding feels like earlier. This, in turn, can allow instructors to focus on teaching judgment and fine-tuning skills rather than foundational technique.


\subsection{Challenges and Future Work}
Our study encountered several challenges arising from both device limitations and procedural design decisions. The AR system currently has controller drift errors, audio trigger latency, and transmission instability caused by hardware performance and network conditions. While recalibration and data cleaning helped mitigate these issues, it did not fully resolve them. Practical usability problems are also present: reduced visual resolution when viewed through the auto-darkening shield and a limited torch rotation range due to the torch attachment impacted user experience. Procedurally, the study design encountered participant fatigue and reduced time for reflection, leading some participants to rush through later tasks. This could be addressed through more flexible lesson scheduling, additional practice opportunities and retry options, longer breaks between sessions, and greater spacing between experimental conditions. Furthermore, situational factors such as occasional equipment glitches, safety checks, and managing the live arc sometimes interrupted the AR learning process. While these challenges did not undermine our findings, they highlight the ongoing difficulties of conducting XR welding research in real-world workshop settings.

Our choice of baseline conditions reflects a balance between validity and experimental control. We examine how AR supports skill acquisition in live welding by addressing constraints inherent to real workshop instruction. Accordingly, we compare the AR system against traditional instructional demonstration. However, as repeating a live demonstration consistently for all participants was logistically impractical, we utilized a video condition to provide a controlled and reproducible baseline. This video replicates standard instructional content, modelled on established practices at our partner welding facility.

A VR-based condition was intentionally excluded due to both safety and methodological concerns. First, a "VR-Live" condition-operating a live welding torch while immersed in VR is inherently unsafe, as restricted environmental awareness poses significant physical risks in a high-heat setting. Second, incorporating a "VR-Training" condition (training in VR followed by a live post-test) would have introduced confounding variables that deviate from our experimental design. Such a setup necessitates a "context switch" between different tracking systems and physical torches, increasing participant burden and session duration. More importantly, in this "VR-Training" condition, it removes critical sensory stressors, specifically intense heat and auditory noise, thereby reducing overall cognitive and physical demands compared to live welding processes. Including a "VR-Training" condition would have fundamentally altered the test conditions, potentially biasing skill acquisition outcomes and making it difficult to isolate the impact of the AR guidance itself. 

Nevertheless, we acknowledge that the absence of a VR benchmark remains a limitation. While our current study demonstrates the benefits of live visual guidance by comparing AR to video-based instructions, a VR comparison could further clarify the role of embodied learning. A direct comparison between our in-situ AR approach and a "VR-Training" system (training in VR followed by a live post-test) using identical visual guidance would be required to discern whether the learning benefits derive solely from the visual overlay or if the embodied sensations of live welding are essential to justifying the technical complexity of in-situ AR.

To address challenges, future systems could integrate higher-precision external tracking solutions, such as hybrid tracking rigs, to provide stable spatial reference points that reduce drift, latency, and instability. This would improve parameter accuracy and ensure more reliable alignment between AR overlays and welding action. The system design could be extended to accommodate a wider range of welding motions and potentially support more complex postures encountered in real-world practice. Furthermore, future work could include the aforementioned comparison with a VR-based training system to better quantify the value of embodied learning. Adaptive learning frameworks could be used to dynamically adjust training settings, including lesson sequences, task complexity, and rest intervals, based on individual performance data. Extended and longer-term studies are also important in order to better capture learning trajectories and skill transfer over time. 

\section{Conclusion}
In this paper, we introduce WeldAR, an in-situ system that overlays real-time instruction and feedback during live MIG welding. Using WeldAR, we investigate how live, in-booth AR guidance can shift welding apprenticeship from post-hoc, inspection-based assessment toward formative and scalable training.

We evaluated WeldAR against video instruction with 24 novices in a within-subjects crossover study. Our findings address our research questions in three ways. First (RQ1), we show that in-situ AR guidance improves novice performance, particularly for work angle and travel speed, and leads to more accurate and stable execution compared to video instruction. Second (RQ2), AR training supports skill transfer: participants maintained performance when feedback was removed and relied on embodied cues, indicating internalization rather than dependence on overlays. Third (RQ3), while AR was perceived as more useful and engaging, interviews revealed ergonomic and cognitive trade-offs, including visual overload and hardware-related fatigue. Together, these findings suggest that AR training systems should move beyond binary correctness toward range-based and gradually reduced feedback that supports embodied learning and safe practice. Next steps for this work include exploring adaptive feedback fading strategies tied to performance stability, improving tracking fidelity and ergonomics for extended use, and conducting longer-term deployments to examine retention and skill transfer over time. Collectively, this work points toward scalable AR training approaches that support embodied skill acquisition in welding and other high-risk vocational domains.

\section*{Acknowledgments}
We would like to thank the Southwest Pennsylvania Build Back Better Initiative, administered through the Block Center for Technology and Society, and the Manufacturing Futures Institute for their support in funding this project. We would also like to thank the Industrial Arts Workshop for their feedback during the system development, and Carnegie Mellon's Tech Spark (John Fulmer and Edward R. Wojciechowski) for their help with the user study. We would also like to extend our thanks to David Lindlbauer and Catarina Goncalves Fidalgo for their valuable feedback and advice on improving the paper.



\bibliographystyle{unsrtnat}
\bibliography{sample-base,Lia_Library,Dina,dina2,ref,references_converted}

@book{polanyi_tacit_2009,
	address = {Chicago ; London},
	edition = {Reissue edition},
	title = {The {Tacit} {Dimension}},
	isbn = {978-0-226-67298-4},
	language = {English},
	publisher = {University Of Chicago Press},
	author = {Polanyi, Michael and Sen, Amartya},
	month = may,
	year = {2009},
}

@inproceedings{chatain_three_2023,
	address = {New York, NY, USA},
	series = {{CHI} {EA} '23},
	title = {Three {Perspectives} on {Embodied} {Learning} in {Virtual} {Reality}: {Opportunities} for {Interaction} {Design}},
	isbn = {978-1-4503-9422-2},
	shorttitle = {Three {Perspectives} on {Embodied} {Learning} in {Virtual} {Reality}},
	url = {https://dl.acm.org/doi/10.1145/3544549.3585805},
	doi = {10.1145/3544549.3585805},
	urldate = {2023-07-23},
	booktitle = {Extended {Abstracts} of the 2023 {CHI} {Conference} on {Human} {Factors} in {Computing} {Systems}},
	publisher = {Association for Computing Machinery},
	author = {Chatain, Julia and Kapur, Manu and Sumner, Robert W.},
	month = apr,
	year = {2023},
	pages = {1--8},
}

@inproceedings{chen_augmenting_2024,
	address = {New York, NY, USA},
	series = {{TEI} '24},
	title = {Augmenting {Embodied} {Learning} in {Welding} {Training}: {The} {Co}-{Design} of an {XR}- and {tinyML}-{Enabled} {Welding} {System} for {Creative} {Arts} and {Manufacturing} {Training}},
	isbn = {9798400704024},
	shorttitle = {Augmenting {Embodied} {Learning} in {Welding} {Training}},
	url = {https://doi.org/10.1145/3623509.3633398},
	doi = {10.1145/3623509.3633398},
	urldate = {2024-01-30},
	booktitle = {Proceedings of the {Eighteenth} {International} {Conference} on {Tangible}, {Embedded}, and {Embodied} {Interaction}},
	publisher = {Association for Computing Machinery},
	author = {Chen, Zhenfang and Johnson, Tate and Knowles, Andrew and Li, Ann and Yi, Semina and Zhuang, Yumeng and Byrne, Daragh and El-Zanfaly, Dina},
	month = feb,
	year = {2024},
	pages = {1--14},
}

@inproceedings{gao_digiclay_2018,
	address = {Hong Kong Hong Kong},
	title = {{DigiClay}: {An} {Interactive} {Installation} for {Virtual} {Pottery} using {Motion} {Sensing} {Technology}},
	isbn = {978-1-4503-6408-9},
	shorttitle = {{DigiClay}},
	url = {https://dl.acm.org/doi/10.1145/3198910.3234659},
	doi = {10.1145/3198910.3234659},
	language = {en},
	urldate = {2024-09-11},
	booktitle = {Proceedings of the 4th {International} {Conference} on {Virtual} {Reality}},
	publisher = {ACM},
	author = {Gao, Zihan and Li, Jingwen and Wang, Huiqiang and Feng, Guangsheng},
	month = feb,
	year = {2018},
	pages = {126--132},
}

@inproceedings{wang_impact_2024,
	address = {Cham},
	title = {The {Impact} of {Feedback} {Mechanism} in {VR} {Learning} {Environment}},
	isbn = {978-3-031-65884-6},
	doi = {10.1007/978-3-031-65884-6_14},
	language = {en},
	booktitle = {Innovative {Technologies} and {Learning}},
	publisher = {Springer Nature Switzerland},
	author = {Wang, Wei-Sheng and Pedaste, Margus and Huang, Yueh-Min},
	editor = {Cheng, Yu-Ping and Pedaste, Margus and Bardone, Emanuele and Huang, Yueh-Min},
	year = {2024},
	pages = {134--142},
}

@article{lassiter2023ai_welding,
  author       = {Lassiter, Tina and Collier, Chelsea and Fleischmann, Kenneth R. and Greenberg, Sherri R.},
  title        = {Welding Instructors' Perspectives on Using AI Technology in Welding Training},
  journal      = {Proceedings of the Association for Information Science and Technology},
  volume       = {60},
  number       = {1},
  pages        = {233--243},
  year         = {2023},
  month        = {October},
  doi          = {10.1002/pra2.784},
}

@article{shankhwar_visuo-haptic_2022,
	title = {A visuo-haptic extended reality–based training system for hands-on manual metal arc welding training},
	volume = {121},
	issn = {1433-3015},
	url = {https://doi.org/10.1007/s00170-022-09328-4},
	doi = {10.1007/s00170-022-09328-4},
	language = {en},
	number = {1},
	urldate = {2022-11-11},
	journal = {The International Journal of Advanced Manufacturing Technology},
	author = {Shankhwar, Kalpana and Chuang, Tung-Jui and Tsai, Yao-Yang and Smith, Shana},
	month = jul,
	year = {2022},
	pages = {249--265},
}

@article{10.1145/3181673,
author = {Hammond, Tracy and Kumar, Shalini Priya Ashok and Runyon, Matthew and Cherian, Josh and Williford, Blake and Keshavabhotla, Swarna and Valentine, Stephanie and Li, Wayne and Linsey, Julie},
title = {It’s Not Just about Accuracy: Metrics That Matter When Modeling Expert Sketching Ability},
year = {2018},
issue_date = {September 2018},
publisher = {Association for Computing Machinery},
address = {New York, NY, USA},
volume = {8},
number = {3},
issn = {2160-6455},
url = {https://doi.org/10.1145/3181673},
doi = {10.1145/3181673},
journal = {ACM Trans. Interact. Intell. Syst.},
month = jul,
articleno = {19},
numpages = {47}
}

@inproceedings{10.1145/3463914.3463921,
author = {Holzwarth, Valentin and Gisler, Joy and Hirt, Christian and Kunz, Andreas},
title = {Comparing the Accuracy and Precision of SteamVR Tracking 2.0 and Oculus Quest 2 in a Room Scale Setup},
year = {2021},
isbn = {9781450389327},
publisher = {Association for Computing Machinery},
address = {New York, NY, USA},
url = {https://doi.org/10.1145/3463914.3463921},
doi = {10.1145/3463914.3463921},
booktitle = {Proceedings of the 2021 5th International Conference on Virtual and Augmented Reality Simulations},
pages = {42–46},
numpages = {5},
location = {Melbourne, VIC, Australia},
series = {ICVARS '21}
}

@Article{act12060257,
AUTHOR = {Pereira, Diogo and Oliveira, Vitor and Vilaça, João L. and Carvalho, Vítor and Duque, Duarte},
TITLE = {Measuring the Precision of the Oculus Quest 2’s Handheld Controllers},
JOURNAL = {Actuators},
VOLUME = {12},
YEAR = {2023},
NUMBER = {6},
ARTICLE-NUMBER = {257},
URL = {https://www.mdpi.com/2076-0825/12/6/257},
ISSN = {2076-0825},
DOI = {10.3390/act12060257}
}

@article{Asplund2020LessonsFT,
  title={Lessons from the welding booth: theories in practice in vocational education},
  author={Stig-b{\"o}rje Asplund and Nina Kilbrink},
  journal={Empirical Research in Vocational Education and Training},
  year={2020},
  volume={12},
  pages={1-23}
}

@article{Fidalgo2023ASO,
  title={A Survey on Remote Assistance and Training in Mixed Reality Environments},
  author={Catarina G. Fidalgo and Yukang Yan and Hyunsung Cho and Maur{\'i}cio Sousa and David Lindlbauer and Joaquim Jorge},
  journal={IEEE Transactions on Visualization and Computer Graphics},
  year={2023},
  volume={29},
  pages={2291-2303}
}

@inproceedings{Uhl2023,
  author    = {Jakob Carl Uhl and Helmut Schrom-Feiertag and Georg Regal and Katja Gallhuber and Manfred Tscheligi},
  title     = {Tangible Immersive Trauma Simulation: Is Mixed Reality the next level of medical skills training?},
  booktitle = {Proceedings of the 2023 CHI Conference on Human Factors in Computing Systems (CHI '23)},
  year      = {2023},
  publisher = {Association for Computing Machinery},
  address   = {New York, NY, USA},
  articleno = {513},
  numpages  = {17},
  doi       = {10.1145/3544548.3581292}
}

@inproceedings{Zhu2025,
  author    = {Yufan Zhu and Ximing Shen and Arata Horie and Yoshihiro Tanaka and Kouta Minamizawa},
  title     = {EmbodyCraft: Exploring Haptic Embodied Experiences for Reflective Practice in Throwing Clay},
  booktitle = {Extended Abstracts of the CHI Conference on Human Factors in Computing Systems (CHI EA '25)},
  year      = {2025},
  publisher = {Association for Computing Machinery},
  address   = {New York, NY, USA},
  articleno = {216},
  numpages  = {7},
  doi       = {10.1145/3706599.3720241}
}

@misc{Seabery2024,
  author       = {Seabery},
  title        = {Soldamatic Augmented Training},
  year         = {2024},
  howpublished = {\url{https://seaberyat.com/en/soldamatic/}},
  note         = {Retrieved September 11, 2025}
}

@misc{Miller2023,
  author       = {{Miller Electric}},
  title        = {AugmentedArc Augmented Reality Welding System},
  year         = {2023},
  howpublished = {\url{https://www.millerwelds.com/-/media/miller-electric/files/pdf/literature-ordering-form/training-solutions/276977_augmentedarc_brochure.pdf}},
  note         = {Retrieved September 11, 2025}
}

@article{AkundiAugmentedRI,
  title={Augmented Reality Integrated Welder Training for Mechanical Engineering Technology},
  author={Aditya Akundi and Hamid Eisazadeh and Monavareh Torabizadeh},
    year         = {2022},
  journal={2022 ASEE Annual Conference \& Exposition Proceedings}
}

@article{Byrd2015TheUO,
  title={The Use of Virtual Welding Simulators to Evaluate Experienced Welders},
  author={Alex Preston Byrd and Ryan G. Anderson and Richard T. Stone},
  journal={Welding Journal},
  year={2015},
  volume={94},
  pages={389-395}
}

@article{Heibel2023VirtualRI,
  title={Virtual reality in welding training and education: A literature review},
  author={Brittney Heibel and Ryan Anderson and Merritt L. Drewery},
  journal={Journal of Agricultural Education},
  year={2023}
}

@article{Deja2024TeachMH,
  title={Teach Me How to ImproVISe: Co-Designing an Augmented Piano Training System for Improvisation},
  author={Jordan Aiko Deja and Sandi Stor and Ilonka Pucihar and Klen Copic Pucihar and Matja{\v{z}} Kljun},
  journal={ArXiv},
  year={2024},
  volume={abs/2402.02999}
}

@inproceedings{Turakhia2025,
  author    = {Dishita Turakhia and Mark Parent and Tovi Grossman and Michael Glueck and Ben Lafreniere},
  title     = {Investigating Augmented Reality for Adaptive Motor-Skill Training},
  booktitle = {Proceedings of Graphics Interface (GI '25)},
  year      = {2025},
  publisher = {Association for Computing Machinery},
  address   = {New York, NY, USA},
  numpages  = {10}
}

@inproceedings{Adiwangsa2025,
  author    = {Michelle Adiwangsa and Penny Sweetser and Anne Ozdowska},
  title     = {Snap, Sweat, and Sketch: Designing Home Exercise Experiences for Augmented Reality Head-mounted Displays},
  booktitle = {Proceedings of the 2025 CHI Conference on Human Factors in Computing Systems (CHI '25)},
  year      = {2025},
  publisher = {Association for Computing Machinery},
  address   = {New York, NY, USA},
  articleno = {211},
  numpages  = {20},
  doi       = {10.1145/3706598.3713575}
}

@inproceedings{Ji2025,
  author    = {Hongfei Ji and Peiyu Hu and Dina El{-}Zanfaly},
  title     = {Reshaping Craft Learning: Insights from Designing an AI-Augmented MR System for Wheel-Throwing},
  booktitle = {Proceedings of the 2025 ACM Designing Interactive Systems Conference (DIS '25)},
  year      = {2025},
  publisher = {Association for Computing Machinery},
  address   = {New York, NY, USA},
  pages     = {2549--2573},
  doi       = {10.1145/3715336.3735844}
}

@article{Yan2021,
  author    = {Wei Yan},
  title     = {Augmented reality applied to LEGO construction: AR-based building instructions with high accuracy \& precision and realistic object-hand occlusions},
  journal   = {Virtual Reality},
  volume    = {25},
  number    = {3},
  pages     = {769--784},
  year      = {2021},
  doi       = {10.1007/s10055-021-00582-7}
}

@article{Johnson2023GeneralizationOP,
  title={Generalization of procedural motor sequence learning after a single practice trial},
  author={B P Johnson and I{\~n}aki Iturrate and Rawan Fakhreddine and Marlene B{\"o}nstrup and Ethan R. Buch and E. M. Robertson and Leonardo G. Cohen},
  journal={NPJ Science of Learning},
  year={2023},
  volume={8}
}

@article{davis1989TAM,
  title={Perceived usefulness, perceived ease of use, and user acceptance of information technology},
  author={Davis, Fred D.},
  journal={MIS Quarterly},
  volume={13},
  number={3},
  pages={319--340},
  year={1989},
  publisher={Management Information Systems Research Center, University of Minnesota},
  doi={10.2307/249008},
  url={https://doi.org/10.2307/249008}
}

@article{Hart1988DevelopmentON,
  title={Development of NASA-TLX (Task Load Index): Results of Empirical and Theoretical Research},
  author={S. G. Hart and Lowell E. Staveland},
  journal={Advances in psychology},
  year={1988},
  volume={52},
  pages={139-183}
}

@article{fereday2006deductive,
  title={Demonstrating rigor using thematic analysis: A hybrid approach of inductive and deductive coding and theme development},
  author={Fereday, Jennifer and Muir-Cochrane, Eimear},
  journal={International Journal of Qualitative Methods},
  volume={5},
  number={1},
  pages={80--92},
  year={2006},
  publisher={SAGE Publications},
  doi={10.1177/160940690600500107},
  url={https://doi.org/10.1177/160940690600500107}
}

@article{klotzbier2025scaffolding,
  title={Scaffolding theory of maturation, cognition, motor performance, and motor skill acquisition: a revised and comprehensive framework for understanding motor-cognitive interactions across the lifespan},
  author={Klotzbier, Thomas J. and Schott, Nadja},
  journal={Frontiers in Human Neuroscience},
  volume={19},
  pages={1531547},
  year={2025},
  publisher={Frontiers},
  doi={10.3389/fnhum.2025.1631958},
  url={https://doi.org/10.3389/fnhum.2025.1631958}
}

@techreport{miller2012om238,
  title       = {Millermatic 212 Auto-Set Owner's Manual},
  author      = {{Miller Electric Mfg. Co.}},
  institution = {Miller Electric Mfg. Co.},
  year        = {2012},
  number      = {OM-238 118F},
  address     = {Appleton, WI}
}

@misc{miller2024migparameters,
  title        = {MIG Welding: Setting the Correct Parameters},
  author       = {{Miller Electric Mfg. Co.}},
  howpublished = {\url{https://www.millerwelds.com/resources/article-library/mig-welding-setting-the-correct-parameters}},
  note         = {Accessed: September 10, 2025},
  year         = {2025}
}

@book{schmidt2011motor,
  author    = {Richard A. Schmidt and Timothy D. Lee},
  title     = {Motor Control and Learning: A Behavioral Emphasis},
  edition   = {5},
  year      = {2011},
  publisher = {Human Kinetics},
  address   = {Champaign, IL}
}

@article{kawato1999internal,
  author    = {Kawato, Mitsuo},
  title     = {Internal models for motor control and trajectory planning},
  journal   = {Current Opinion in Neurobiology},
  volume    = {9},
  number    = {6},
  pages     = {718--727},
  year      = {1999},
  publisher = {Elsevier},
  doi       = {10.1016/S0959-4388(99)00028-8}
}

@article{aoyagi2019fading,
  author  = {Aoyagi, Yoichiro and Ohnishi, Eri and Yamamoto, Yoshinori and Kado, Naoki and Suzuki, Toshiaki and Ohnishi, Hitoshi and Hokimoto, Nozomi and Fukaya, Naomi},
  title   = {Feedback protocol of `fading knowledge of results' is effective for prolonging motor learning retention},
  journal = {Journal of Physical Therapy Science},
  volume  = {31},
  number  = {8},
  pages   = {687--691},
  year    = {2019},
  doi     = {10.1589/jpts.31.687}
}

@article{Hadjimichael2024TacitEmbodiment,
  author       = {Demetris Hadjimichael and Rodrigo Ribeiro and Haridimos Tsoukas},
  title        = {How Does Embodiment Enable the Acquisition of Tacit Knowledge in Organizations? From Polanyi to Merleau-Ponty},
  journal      = {Organization Studies},
  volume       = {45},
  number       = {4},
  pages        = {545--570},
  year         = {2024},
  doi          = {10.1177/01708406241228374}
}

@article{Zenner2021HaRTT,
  title={HaRT - The Virtual Reality Hand Redirection Toolkit},
  author={Andr{\'e} Zenner and Hannah Maria Kriegler and Antonio Kr{\"u}ger},
  journal={Extended Abstracts of the 2021 CHI Conference on Human Factors in Computing Systems},
  year={2021}
}

@article{Bschel2021MIRIAAM,
  title={MIRIA: A Mixed Reality Toolkit for the In-Situ Visualization and Analysis of Spatio-Temporal Interaction Data},
  author={Wolfgang B{\"u}schel and Anke Lehmann and Raimund Dachselt},
  journal={Proceedings of the 2021 CHI Conference on Human Factors in Computing Systems},
  year={2021}
}

@article{jensen2018vr,
  author    = {Jensen, Lasse and Konradsen, Flemming},
  title     = {A review of the use of virtual reality head-mounted displays in education and training},
  journal   = {Education and Information Technologies},
  year      = {2018},
  volume    = {23},
  pages     = {1515--1529},
  doi       = {10.1007/s10639-017-9676-0}
}

@article{xu2021hmd,
  author    = {Xu, Xuanhui and Mangina, Eleni and Campbell, Abraham G},
  title     = {HMD-based virtual and augmented reality in medical education: a systematic review},
  journal   = {Frontiers in Virtual Reality},
  year      = {2021},
  volume    = {2},
  pages     = {692103},
  doi       = {10.3389/frvir.2021.692103}
}

@misc{weldvrWeldingSimulator,
	author = {WELDVR},
	title = {{W}elding {S}imulator --- weldvr.com},
	howpublished = {\url{https://weldvr.com/}},
	year = {2025},
	note = {[Accessed 05-12-2025]},
}

@INPROCEEDINGS{Ipsita2022-lq,
  title      = "Towards modeling of virtual reality welding simulators to
                promote accessible and scalable training",
  booktitle  = "{CHI} Conference on Human Factors in Computing Systems",
  author     = "Ipsita, Ananya and Erickson, Levi and Dong, Yangzi and Huang,
                Joey and Bushinski, Alexa K and Saradhi, Sraven and Villanueva,
                Ana M and Peppler, Kylie A and Redick, Thomas S and Ramani,
                Karthik",
  publisher  = "ACM",
  month      =  apr,
  year       =  2022,
  address    = "New York, NY, USA",
  conference = "CHI '22: CHI Conference on Human Factors in Computing Systems",
  location   = "New Orleans LA USA"
}

@ARTICLE{Lee2023-ca,
  title     = "A multisensor interface to improve the learning experience in
               arc welding training tasks",
  author    = "Lee, Hoi-Yin and Zhou, Peng and Duan, Anqing and Wang, Jiangliu
               and Wu, Victor and Navarro-Alarcon, David",
  journal   = "IEEE Trans. Hum. Mach. Syst.",
  publisher = "Institute of Electrical and Electronics Engineers (IEEE)",
  volume    =  53,
  number    =  3,
  pages     = "619--628",
  month     =  jun,
  year      =  2023,
  copyright = "https://ieeexplore.ieee.org/Xplorehelp/downloads/license-information/IEEE.html"
}

@misc{metaConsiderations,
	author = {Meta},
	title = {{K}ey considerations --- developers.meta.com},
	howpublished = {\url{https://developers.meta.com/horizon/design/mr-design-guideline/}},
	year = {2025},
	note = {[Accessed 05-12-2025]},
}

@article{Lassiter2023WeldingIP,
  title={Welding Instructors' Perspectives on Using AI Technology in Welding Training},
  author={Tina B. Lassiter and Chelsea Collier and Kenneth R. Fleischmann and Sherri R. Greenberg},
  journal={Proceedings of the Association for Information Science and Technology},
  year={2023},
  volume={60}
}

@article{Huang2021AdapTutAR,
  title={AdapTutAR: An Adaptive Tutoring System for Machine Tasks in Augmented Reality},
  author={Gaoping Huang and Xun Qian and Tianyi Wang and Fagun Patel and Maitreya Sreeram and Yuanzhi Cao and Karthik Ramani and Alexander J. Quinn},
  journal={Proceedings of the 2021 CHI Conference on Human Factors in Computing Systems},
  year={2021}
}

@article{Chan2022VRAA,
  title={VR and AR virtual welding for psychomotor skills: a systematic review},
  author={Vei Siang Chan and Habibah Norehan Hj Haron and Muhammad Ismail Mat Isham and Farhan bin Mohamed},
  journal={Multimedia Tools and Applications},
  year={2022},
  volume={81},
  pages={12459 - 12493}
}

@inproceedings{OrigamiSensei,
author = {Chen, Qiyu and Mishra, Richa and Purnamasari, Lia Sparingga and EL-Zanfaly, Dina and Kitani, Kris},
title = {Origami Sensei: A Mixed Reality AI-Assistant},
year = {2025},
isbn = {9798400713941},
publisher = {Association for Computing Machinery},
address = {New York, NY, USA},
url = {https://doi.org/10.1145/3706598.3714099},
doi = {10.1145/3706598.3714099},
booktitle = {Proceedings of the 2025 CHI Conference on Human Factors in Computing Systems},
articleno = {1127},
numpages = {18},
location = {
},
series = {CHI '25}
}

@inproceedings{AugmentingWeldingTraining,
author = {Johnson, Tate and Li, Ann and Knowles, Andrew and Chen, Zhenfang and Yi, Semina and Zhuang, Yumeng and El-Zanfaly, Dina and Byrne, Daragh},
title = {Augmenting Welding Training: An XR Platform to Foster Muscle Memory and Mindfulness for Skills Development},
year = {2023},
isbn = {9798400704253},
publisher = {Association for Computing Machinery},
address = {New York, NY, USA},
url = {https://doi.org/10.1145/3626485.3626544},
doi = {10.1145/3626485.3626544},
booktitle = {Companion Proceedings of the 2023 Conference on Interactive Surfaces and Spaces},
pages = {61–64},
numpages = {4},
location = {Pittsburgh, PA, USA},
series = {ISS Companion '23}
}

@article{wang_critical_2025,
	title = {Critical {Anatomy}-{Preserving} and {Terrain}-{Augmenting} {Navigation} ({CAPTAiN}): {Application} to {Laminectomy} {Surgical} {Education}},
	volume = {7},
	issn = {2576-3202},
	shorttitle = {Critical {Anatomy}-{Preserving} and {Terrain}-{Augmenting} {Navigation} ({CAPTAiN})},
	url = {https://ieeexplore.ieee.org/document/11082416},
	doi = {10.1109/TMRB.2025.3589795},
	number = {3},
	urldate = {2026-02-15},
	journal = {IEEE Transactions on Medical Robotics and Bionics},
	author = {Wang, Jonathan and Ishida, Hisashi and Usevitch, David and Venkatesh, Kesavan and Wang, Yi and Armand, Mehran and Bronheim, Rachel and Jain, Amit and Munawar, Adnan},
	month = aug,
	year = {2025},
	pages = {1125--1138},
}

@article{chan_use_2026,
	title = {The {Use} of {Immersive} {Virtual} {Reality} {Surgical} {Simulation} to {Increase} {Medical} {Student} {Confidence} and {Surgical} {Knowledge} in {Orthopedics}},
	volume = {83},
	issn = {1931-7204},
	url = {https://www.sciencedirect.com/science/article/pii/S1931720425003721},
	doi = {10.1016/j.jsurg.2025.103791},
	number = {1},
	urldate = {2026-02-15},
	journal = {Journal of Surgical Education},
	author = {Chan, Elizabeth and Movassaghi, Aghdas and Smith, Lana and McKinley, Matthew and Osswald, Roya and Lubert, Jocelyn and Sabesan, Vani J.},
	month = jan,
	year = {2026},
	pages = {103791},
}

@article{stone_autonomous_2025,
	title = {Autonomous {Educational} {System} for {Surgical} {Training} {Utilizing} {Deep} {Learning} {Combined} with {Extended} {Reality}},
	volume = {2},
	issn = {2994-1520},
	url = {https://journals.sagepub.com/action/showAbstract},
	doi = {10.1177/29941520251361898},
	language = {EN},
	number = {1},
	urldate = {2026-02-15},
	journal = {Journal of Medical Extended Reality},
	publisher = {SAGE Publications},
	author = {Stone, Jonathan J. and Stone, Nelson N. and Griffith, Steven H. and Zeller, Kyle and Wilson, Michael P.},
	month = mar,
	year = {2025},
	pages = {29941520251361898},
}

@article{preston_byrd_dexterity_2018,
	title = {Dexterity: {An} {Indicator} of {Future} {Performance} in {Beginning} {Welders}},
	volume = {43},
	shorttitle = {Dexterity},
	doi = {10.5328/cter43.2.195},
	number = {2},
	journal = {Career and Technical Education Research},
	author = {Preston Byrd, A. and Stone, Richard T. and Anderson, Ryan G.},
	month = sep,
	year = {2018},
	pages = {195--212},
}

@article{huang_research_2020,
	title = {Research on {Teaching} a {Welding} {Implementation} {Course} {Assisted} by {Sustainable} {Virtual} {Reality} {Technology}},
	volume = {12},
	copyright = {http://creativecommons.org/licenses/by/3.0/},
	issn = {2071-1050},
	url = {https://www.mdpi.com/2071-1050/12/23/10044},
	doi = {10.3390/su122310044},
	language = {en},
	number = {23},
	urldate = {2026-02-16},
	journal = {Sustainability},
	publisher = {publisher},
	author = {Huang, Chuang-Yeh and Lou, Shi-Jer and Cheng, Yuh-Ming and Chung, Chih-Chao},
	month = nov,
	year = {2020},
}

@article{liang_simple_2014,
	title = {Simple {Virtual} {Reality} {Skill} {Training} {System} for {Manual} {Arc} {Welding}},
	volume = {26},
	url = {https://www.fujipress.jp/jrm/rb/robot002600010078/},
	doi = {10.20965/jrm.2014.p0078},
	number = {1},
	urldate = {2026-02-16},
	journal = {Journal of Robotics and Mechatronics},
	publisher = {Fuji Technology Press Ltd.},
	author = {Liang, Xin and Kato, Hideo and Hashimoto, Nobuyoshi and Okawa, Kazuya},
	month = feb,
	year = {2014},
	pages = {78--84},
}

@article{wells_effect_2020,
	title = {The {Effect} of {Virtual} {Reality} {Technology} on {Welding} {Skill} {Performance}},
	volume = {61},
	copyright = {Copyright (c)},
	issn = {1042-0541},
	url = {https://jae-online.org/index.php/jae/article/view/2054},
	doi = {10.5032/jae.2020.01152},
	language = {en},
	number = {1},
	urldate = {2026-02-16},
	journal = {Journal of Agricultural Education},
	author = {Wells, Trent and Miller, Greg},
	month = mar,
	year = {2020},
	pages = {152--171},
}

@article{shankhwar_interactive_2022,
	title = {An interactive extended reality-based tutorial system for fundamental manual metal arc welding training},
	volume = {26},
	issn = {1434-9957},
	url = {https://doi.org/10.1007/s10055-022-00626-6},
	doi = {10.1007/s10055-022-00626-6},
	language = {en},
	number = {3},
	urldate = {2026-02-16},
	journal = {Virtual Reality},
	author = {Shankhwar, Kalpana and Smith, Shana},
	month = sep,
	year = {2022},
	pages = {1173--1192},
}

@article{chung_research_2020,
	title = {Research on {Optimization} of {VR} {Welding} {Course} {Development} with {ANP} and {Satisfaction} {Evaluation}},
	volume = {9},
	copyright = {http://creativecommons.org/licenses/by/3.0/},
	issn = {2079-9292},
	url = {https://www.mdpi.com/2079-9292/9/10/1673},
	doi = {10.3390/electronics9101673},
	language = {en},
	number = {10},
	urldate = {2026-02-16},
	journal = {Electronics},
	publisher = {publisher},
	author = {Chung, Chih-chao and Tung, Chun-Chun and Lou, Shi-Jer},
	month = oct,
	year = {2020},
}

@article{rodriguez-martin_learning_2019,
	title = {Learning methodology based on weld virtual models in the mechanical engineering classroom},
	volume = {27},
	copyright = {© 2019 Wiley Periodicals, Inc.},
	issn = {1099-0542},
	url = {https://onlinelibrary.wiley.com/doi/abs/10.1002/cae.22140},
	doi = {10.1002/cae.22140},
	language = {en},
	number = {5},
	urldate = {2026-02-16},
	journal = {Computer Applications in Engineering Education},
	author = {Rodríguez-Martín, Manuel and Rodríguez-Gonzálvez, Pablo},
	year = {2019},
	note = {\_eprint: https://onlinelibrary.wiley.com/doi/pdf/10.1002/cae.22140},
	pages = {1113--1125},
}

@article{Yu2024DesignSO,
  title={Design Space of Visual Feedforward And Corrective Feedback in XR-Based Motion Guidance Systems},
  author={Xingyao Yu and Benjamin Lee and Michael Sedlmair},
  journal={Proceedings of the 2024 CHI Conference on Human Factors in Computing Systems},
  year={2024}
}

@String{Computing = "Computing" }

@String{Computer = "{IEEE} Computer" }

@String{Chelsea = "Chelsea" }

@String{Springer = "Springer-Verlag" }

@article{article,
    author = {Burghouts, Gertjan and Geusebroek, Jan-Mark},
    year = {2009},
    month = {February},
    pages = {306-313},
    title = {Material-specific adaptation of color invariant features},
    volume = {30},
    journal = {Pattern Recognition Letters},
    doi = {10.1016/j.patrec.2008.10.005}
}

\clearpage
\appendix
\section{System Validation of In-the-Wild Deployment}
\label{app:system_validation}
\subsection{Purpose.}

\begin{figure}[!htbp]
\centering
\includegraphics[width=1\columnwidth, alt={Two setups for a simulated welding experiment. Left: a welding torch and protective helmet are mounted on stands, both fixed in place. Right: a person wearing gloves holds the torch by hand, while the helmet remains fixed on a stand.}]{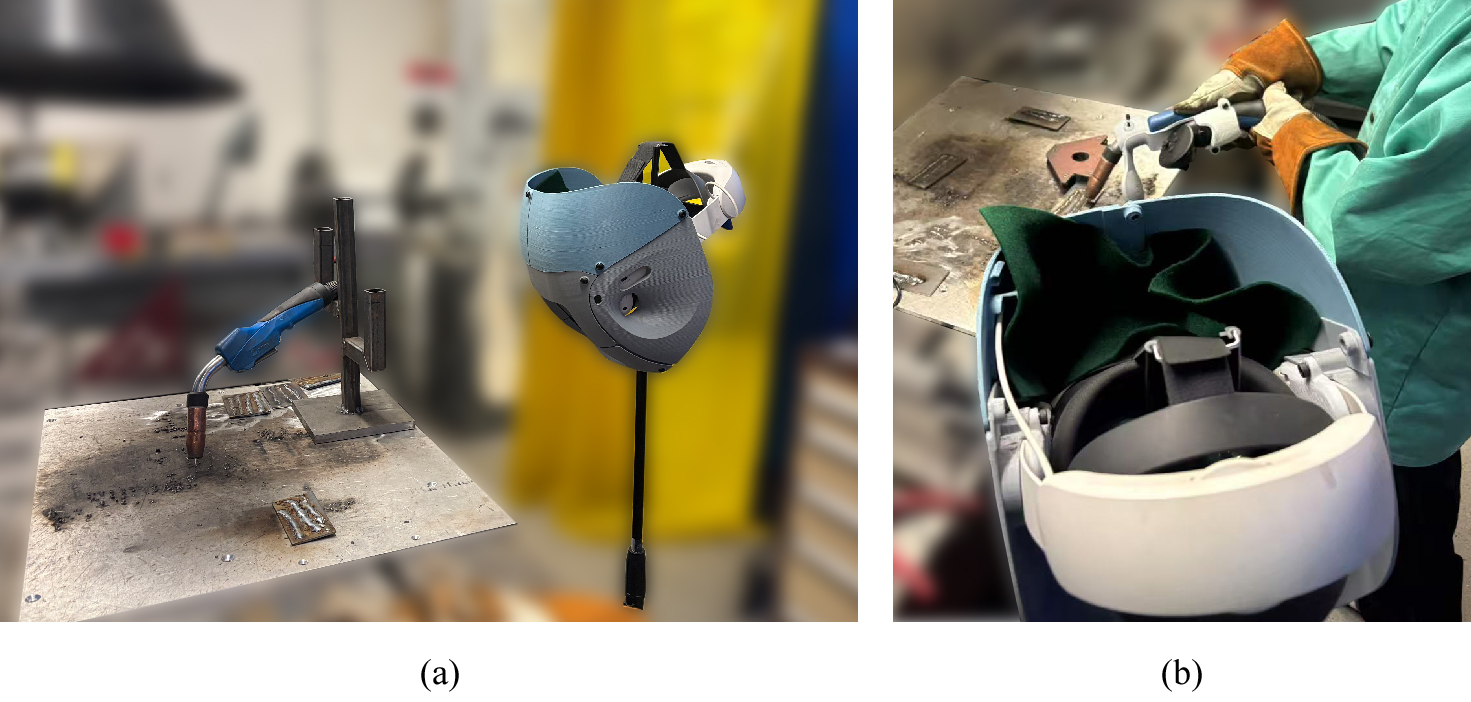}
\caption{Controller drift experiment.\textbf{ (1)} torch and helmet fixed in position. \textbf{(2)} helmet fixed, torch handheld.}
\label{fig:onsite}
\end{figure}

\revised{To address tracking instability and documented controller errors, specifically headset fluctuations ($M = 0.001$ m, $SD = 0.004$ m) \citep{10.1145/3463914.3463921} and translational errors ranging from 1.24 to 13.75 mm \citep{act12060257}, we conducted validation tests to mitigate drift caused by welding sparks and the auto-darkening screen. We utilized Meta Quest Pro controllers in the study environment, employing simulated welding with audio cues for consistency. The protocol mirrored the user study, enforcing initial workspace setup and per-line torch calibration (Figure~\ref{fig:onsite}).}

\subsection{Experiments}

\paragraph{Controller drift.} To evaluate controller drift, \revised{we evaluated controller drift through 40 simulated weld trials (30s each) varying lighting, torch handling, and screen states. We quantified drift by calculating the proportion of affected frames in the handheld/screen-on condition and estimated worst-case line-level probabilities via 10,000 bootstrap samples.}

\begin{figure}[!htbp]
\centering
\includegraphics[width=\columnwidth, alt={Sequence of welding skill parameter setup. Left: a gloved hand uses a digital caliper to measure the contact tip to work distance for determining travel speed. Middle: a welding torch is secured at a 45-degree travel angle using a triangular magnetic weld clamp. Right: the welding torch is repositioned and clamped at a 45-degree work angle on the same magnetic holder.}]{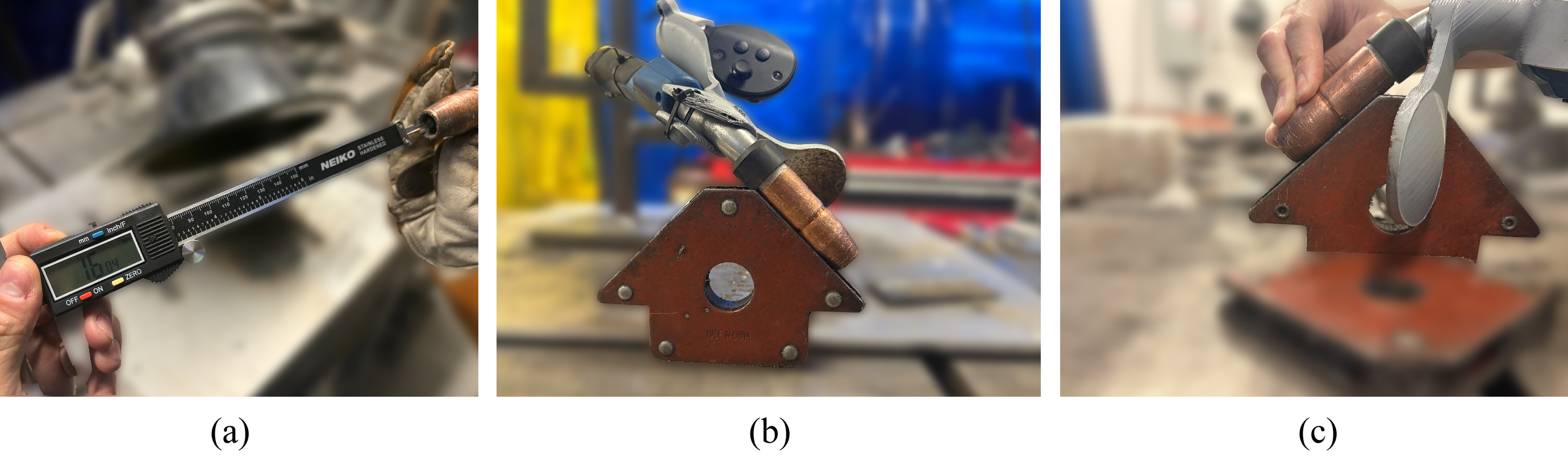}
\caption{Skill parameter measurements. \textbf{(a)} Measuring length for CTWD and travel speed. \textbf{(b)} Fixing torch to 45° travel angle on a magnetic weld clamp. \textbf{(c)} Fixing torch to 45° work angle on a magnetic weld clamp.}
\label{fig:parameter}
\end{figure}

\paragraph{Skill parameter accuracy.} \revised{We validated skill parameters by comparing system logs against physical ground truths: measured wire lengths for CTWD, precision jigs ($30^{\circ}, 45^{\circ}, 60^{\circ}$) for angles, and timed displacement over fixed distances for travel speed (see Figure~\ref{fig:parameter}).}

\paragraph{Microphone latency.} \revised{To measure audio latency, we recorded detection delays for 5-second weld sounds played on the worktable across 50 trials per condition. We manipulated buffer size (128 vs. 1024) and microphone distance (0.5 m vs. 2 m) to identify optimal settings.}

\subsection{Results}
\paragraph{Controller drift.} \revised{Drift primarily affected CTWD and speed in low-light, handheld conditions (~7\% frame error; ~25\% line error). While bootstrap analysis indicated a 50\% drift probability for extended sequences, these specific characteristics confirmed that drift is manageable through frequent recalibration.}

\paragraph{Skill parameter accuracy.} \revised{The system demonstrated high stability against ground truth. Angular errors were minimal ($<1^{\circ}$), and displacement-based metrics remained within usable bounds for training (CTWD error $M \approx 2.6$ mm; Speed deviation $<1.1$ IPM), confirming sufficient accuracy for instruction.}

\paragraph{Microphone latency.} \revised{Buffer size proved the dominant factor for stability. A 128-frame buffer yielded consistent, low-latency performance ($M \approx 0.20$ s, $SD = 0.006$) independent of distance, establishing a reliable trigger for data collection.}

\revised{\subsection{Mitigation Strategies} To ensure data validity despite hardware limitations, we implemented four specific measures:}

\begin{itemize}
    \item \textbf{Environment.} We arranged additional spotlights to enhance illumination for both helmet and controllers, and we positioned the microphone at 0.5 m from the weld table with a 128-frame buffer.
    \item \textbf{Drift screening.} We excluded lines with CTWD $<$ 0 and flagged extreme initial CTWD values as invalid, based on comparisons to actual weld artifacts.
    \revised{\item \textbf{Procedural Redundancy:} Enforced per-line recalibration and doubled trial counts to maintain statistical power after excluding drift-affected lines.}
    \item \textbf{Data alignment.} We shifted log windows by 20 frames($~280ms$ under 90 Hz) post-weld to offset audio delay. Together, these measures enabled reliable data collection during the user study.

\end{itemize}

\revised{These steps effectively minimized the impact of measurement artifacts, ensuring that subsequent analyses focused on learning outcomes, though we acknowledge minor residual drift remains.}

\clearpage
\onecolumn
\section{Survey}
\label{app:survey}
\begin{table}[ht]
\centering
\caption{Survey questions with response scales for AR and Video conditions}
\begin{tabular}{p{0.55\textwidth} p{0.20\textwidth} p{0.20\textwidth}}
\hline
\textbf{Question} & \textbf{AR Rating} & \textbf{Video Rating} \\
\hline
The training tool made it easier to learn how to weld. & Strongly Disagree -- Strongly Agree & Strongly Disagree -- Strongly Agree \\
The training tool helped me understand basic welding techniques. & Strongly Disagree -- Strongly Agree & Strongly Disagree -- Strongly Agree \\
The training tool helped me complete my first welding task successfully. & Strongly Disagree -- Strongly Agree & Strongly Disagree -- Strongly Agree \\
I believe the training tool is helpful for beginners learning to weld. & Strongly Disagree -- Strongly Agree & Strongly Disagree -- Strongly Agree \\
The training tool was easy to use. & Strongly Disagree -- Strongly Agree & Strongly Disagree -- Strongly Agree \\
I was able to understand how to use the tool without much help. & Strongly Disagree -- Strongly Agree & Strongly Disagree -- Strongly Agree \\
The training tool makes the learning process engaging and enjoyable. & Strongly Disagree -- Strongly Agree & Strongly Disagree -- Strongly Agree \\
I feel confident in applying the skills I learned through the training tool. & Strongly Disagree -- Strongly Agree & Strongly Disagree -- Strongly Agree \\
I would like to use the training tool again to improve my welding skills. & Strongly Disagree -- Strongly Agree & Strongly Disagree -- Strongly Agree \\
I would recommend the training tool to others. & Strongly Disagree -- Strongly Agree & Strongly Disagree -- Strongly Agree \\
How mentally demanding was it to use the training tool? & Very High -- Very Low & Very High -- Very Low \\
How physically demanding was it to use the training tool? & Very High -- Very Low & Very High -- Very Low \\
How fast-paced or hurried was the training experience? & Very Rushed -- Very Relaxed & Very Rushed -- Very Relaxed \\
How well were you able to complete the welding task with the help of the training tool? & Failure -- Excellent Performance & Failure -- Excellent Performance \\
How much effort did it take to learn welding using the training tool? & Very High -- Very Low & Very High -- Very Low \\
How frustrated did you feel while completing the task? & Very High -- Very Low & Very High -- Very Low \\
\hline
\end{tabular}
\label{tab:survey_questions_simple}
\end{table}

\clearpage
\section{Questionnaire}
\label{app:questionnaire}

\subsubsection*{1. General Experience}
\begin{enumerate}
  \item Can you describe your overall experience with both training methods during the session?
  \item Can you describe your general idea about welding before and after the study, and how each session influenced your perception of welding?
  \item Right handed/left handed
\end{enumerate}

\subsubsection*{2. Experience During the Session}
\begin{enumerate}
  \item How would you describe your experience using the augmented interfaces (UI) in AR compared to the video training?
  \item How do you pay attention to the AR feedback?
  \item Scoring
  \item What are your thoughts on the way the learning modules were structured in each method? What aspects worked well and what aspects did not work for you?
  \item Based on your scores, what aspects of each training method were easy or difficult to use? Did you encounter any challenges or frustrations in either method?
  \item In each method, was there a moment when you felt supported during the training?
  \item In each method, was there a moment when you felt distracted during the training?
  \item Based on your scores, how would you describe your engagement with each training method, and what influenced your experience?
  \item Based on your scores, how would you describe your experience using the equipment (helmet, torch) in AR compared to the setup for the video training?
\end{enumerate}

\subsubsection*{3. Impact on Performance}
\begin{enumerate}
  \item Based on your scores, how did each training method affect your welding performance?
  \item How did each method influence the way you completed your tasks? In which method did you feel more able to remember the techniques?
  \item Based on your scores, how did each training method affect your confidence in welding?
\end{enumerate}

\subsubsection*{4. Reflection and Evaluation}
\begin{enumerate}
  \item Between the two training methods, which one was most useful to you? Why?
  \item Based on your scores, what factors would influence your decision to use one of these training methods again in the future?
  \item Based on your scores, how likely would you be to recommend each training method to someone learning welding? Why?
\end{enumerate}

\subsubsection*{5. Suggestions}
\begin{enumerate}
  \item Do you have any other suggestions?
\end{enumerate}

\section{Performance Data}
\label{app:performance_data}
\vspace*{-\baselineskip} 

\begin{table*}[t!]
  \caption{\revised{Summary of means (M) and standard deviations (SD) for z-scored deviation and variability across composite and skill-level measures. Note that Z-scores were calculated separately for the Assisted pool and the Unassisted pool to allow for within-context comparison. Lower scores indicate better performance.}}
  \label{tab:perf_summary}
  \centering
  \renewcommand{\arraystretch}{0.75}
  \begin{tabular}{lll cccc @{\hskip 0.3in} cccc}
    \toprule
    & & & \multicolumn{4}{c}{\textbf{Combination Session (Assisted)}} & \multicolumn{4}{c}{\textbf{Test Session (Unassisted)}} \\
    \cmidrule(r){4-7} \cmidrule(l){8-11}
    \textbf{Seq.} & \textbf{Cond.} & \textbf{Measure} &
    \multicolumn{2}{c}{\textbf{Deviation}} & \multicolumn{2}{c}{\textbf{Variability}} &
    \multicolumn{2}{c}{\textbf{Deviation}} & \multicolumn{2}{c}{\textbf{Variability}} \\
    \cmidrule(r){4-5} \cmidrule(lr){6-7} \cmidrule(lr){8-9} \cmidrule(l){10-11}
    & & & \textbf{M} & \textbf{SD} & \textbf{M} & \textbf{SD} & \textbf{M} & \textbf{SD} & \textbf{M} & \textbf{SD} \\
    \midrule
    \textbf{AR-first} & AR & Composite & -0.20 & 0.31 & 0.22 & 0.25 & -0.13 & 0.31 & 0.22 & 0.14 \\
     & & CTWD & 0.07 & 0.92 & 0.58 & 0.85 & 0.14 & 0.70 & 0.63 & 0.42 \\
     & & Travel angle & -0.26 & 0.04 & 0.04 & 0.10 & -0.23 & 0.15 & 0.07 & 0.17 \\
     & & Work angle & -0.34 & 0.03 & 0.03 & 0.06 & -0.22 & 0.46 & 0.17 & 0.25 \\
     & & Speed & -0.26 & 0.70 & 0.39 & 0.42 & -0.22 & 0.60 & 0.44 & 0.38 \\
    \addlinespace
    \textbf{AR-first} & Video & Composite & -0.08 & 0.29 & 0.22 & 0.18 & -0.20 & 0.28 & 0.20 & 0.09 \\
     & & CTWD & -0.04 & 0.52 & 0.51 & 0.63 & -0.16 & 0.36 & 0.60 & 0.40 \\
     & & Travel angle & 0.04 & 0.45 & 0.28 & 0.33 & -0.06 & 0.35 & 0.13 & 0.17 \\
     & & Work angle & -0.17 & 0.36 & 0.15 & 0.21 & -0.28 & 0.38 & 0.15 & 0.22 \\
     & & Speed & -0.14 & 0.76 & 0.32 & 0.23 & -0.29 & 0.87 & 0.22 & 0.21 \\
    \midrule
    \textbf{Video-first} & AR & Composite & -0.36 & 0.07 & 0.10 & 0.09 & -0.25 & 0.24 & 0.20 & 0.09 \\
     & & CTWD & -0.26 & 0.19 & 0.33 & 0.36 & -0.14 & 0.23 & 0.72 & 0.34 \\
     & & Travel angle & -0.27 & 0.02 & 0.01 & 0.03 & -0.20 & 0.17 & 0.09 & 0.20 \\
     & & Work angle & -0.35 & 0.01 & 0.01 & 0.01 & -0.33 & 0.27 & 0.16 & 0.24 \\
     & & Speed & -0.59 & 0.14 & 0.17 & 0.17 & -0.34 & 0.55 & 0.27 & 0.20 \\
    \addlinespace
    \textbf{Video-first} & Video & Composite & 0.65 & 0.99 & 0.35 & 0.29 & 0.63 & 1.13 & 0.31 & 0.27 \\
     & & CTWD & 0.23 & 0.50 & 0.77 & 0.56 & 0.21 & 0.75 & 0.90 & 1.03 \\
     & & Travel angle & 0.50 & 1.55 & 0.47 & 0.88 & 0.57 & 1.96 & 0.19 & 0.28 \\
     & & Work angle & 0.82 & 1.55 & 0.56 & 0.48 & 0.88 & 1.61 & 0.44 & 0.26 \\
     & & Speed & 1.03 & 0.86 & 0.66 & 0.25 & 0.85 & 1.03 & 0.57 & 0.24 \\
    \bottomrule
  \end{tabular}
\end{table*}

\begin{table*}[t!]
\centering
\renewcommand{\arraystretch}{0.75}
\caption{\revised{Z-scored deviation by sequence, condition, and segment}}
\begin{tabular}{llrrrrrrrr}
\toprule
Sequence & Condition 
& \makecell{CTWD} 
& \makecell{Travel\\Angle} 
& \makecell{Work\\Angle} 
& \makecell{Travel\\Speed} 
& \makecell{Comb.~1} 
& \makecell{Comb.~2} 
& \makecell{Comb.~3} 
& Test \\
\midrule
AR-first    & AR    &  0.952 &  0.769 &  0.294 &  0.173 & -0.319 & -0.429 & -0.351 & -0.286 \\
AR-first    & Video & -0.260 & -0.204 & -0.238 & -0.288 & -0.260 & -0.242 & -0.276 & -0.296 \\
Video-first & AR    & -0.010 &  0.058 & -0.033 & -0.225 & -0.479 & -0.504 & -0.513 & -0.355 \\
Video-first & Video &  0.991 &  0.651 &  0.334 &  0.343 &  0.382 &  0.456 &  0.185 &  0.362 \\
\bottomrule
\end{tabular}
\label{tab:zscored_deviation}
\end{table*}

\begin{table*}[t!]
\centering
\caption{\revised{Z-scored deviation per participant, sequence, condition, and segment}}
\renewcommand{\arraystretch}{0.8}
\resizebox{\textwidth}{!}{%
\begin{tabular}{lllrrrrrrrr}
\toprule
Participant & Sequence & Condition 
& \makecell{CTWD} 
& \makecell{Travel\\Angle} 
& \makecell{Work\\Angle} 
& \makecell{Travel\\Speed} 
& \makecell{Comb.~1} 
& \makecell{Comb.~2} 
& \makecell{Comb.~3} 
& Test \\
\midrule
3  & Video-first & AR    & -0.434 & -0.337 & -0.390 & -0.474 & -0.571 & -0.567 & -0.565 & -0.549 \\
3  & Video-first & Video & -0.278 & -0.046 & -0.284 & -0.377 & -0.435 & -0.497 & -0.425 & -0.469 \\
5  & Video-first & AR    &  0.159 &  0.474 & -0.078 & -0.273 & -0.318 & -0.466 & -0.462 & -0.064 \\
5  & Video-first & Video &  1.653 &  2.881 &  1.162 &  0.732 &  0.815 &  1.261 &  0.417 &  0.971 \\
7  & Video-first & AR    & -0.416 &  0.019 & -0.250 & -0.389 & -0.522 & -0.512 & -0.476 & -0.514 \\
7  & Video-first & Video & -0.093 & -0.326 & -0.168 & -0.191 & -0.462 & -0.114 & -0.194 & -0.128 \\
9  & Video-first & AR    & -0.235 &  0.013 & -0.092 & -0.171 & -0.476 & -0.523 & -0.544 & -0.396 \\
9  & Video-first & Video &  1.810 &  0.923 &  0.439 &  0.082 & -0.044 &  0.020 & -0.090 & -0.046 \\
11 & Video-first & AR    & -0.137 & -0.069 & -0.081 & -0.361 & -0.505 & -0.522 & -0.560 & -0.457 \\
11 & Video-first & Video &  0.421 & -0.027 & -0.022 & -0.103 & -0.037 &  0.270 &  0.243 &  0.012 \\
13 & Video-first & AR    &  0.254 &  0.193 & -0.230 & -0.352 & -0.329 & -0.485 & -0.441 & -0.432 \\
13 & Video-first & Video &  1.548 &  0.709 &  0.095 &  0.217 &  0.128 &  0.230 & -0.117 & -0.017 \\
15 & Video-first & AR    & -0.451 &  0.127 & -0.014 &  0.080 & -0.496 & -0.556 & -0.550 & -0.496 \\
15 & Video-first & Video &  1.008 &  0.021 &  0.303 & -0.105 &  0.319 &  0.175 & -0.014 &  0.203 \\
17 & Video-first & AR    & -0.059 & -0.139 &  0.103 & -0.460 & -0.500 & -0.514 & -0.513 & -0.387 \\
17 & Video-first & Video &  0.953 &  0.137 & -0.120 &  0.032 & -0.100 & -0.035 &  0.056 & -0.088 \\
19 & Video-first & AR    & -0.288 & -0.234 & -0.112 & -0.437 & -0.558 & -0.525 & -0.513 & -0.282 \\
19 & Video-first & Video &  0.368 & -0.053 &  0.005 &  0.051 &  0.179 & -0.006 & -0.114 & -0.089 \\
21 & Video-first & AR    & -0.084 & -0.084 & -0.181 & -0.527 & -0.544 & -0.561 & -0.484 & -0.502 \\
21 & Video-first & Video &  1.060 &  0.471 &  0.111 &  0.095 & -0.102 &  0.228 & -0.025 & -0.180 \\
23 & Video-first & AR    &  0.816 &  0.626 &  0.263 & -0.246 & -0.389 & -0.275 & -0.505 &  0.013 \\
23 & Video-first & Video &  1.499 &  1.086 &  0.811 &  1.063 &  0.897 &  0.786 &  0.926 &  0.963 \\
26 & Video-first & AR    &  0.754 &  0.108 &  0.671 &  0.911 & -0.541 & -0.539 & -0.545 & -0.199 \\
26 & Video-first & Video &  1.946 &  2.037 &  1.679 &  2.618 &  3.429 &  3.158 &  1.559 &  3.207 \\
\midrule
1  & AR-first & AR    &  1.990 &  0.386 &  0.067 &  0.193 & -0.305 & -0.530 & -0.504 & -0.392 \\
1  & AR-first & Video & -0.310 & -0.434 & -0.430 & -0.495 & -0.400 & -0.419 & -0.477 & -0.455 \\
4  & AR-first & AR    &  1.457 &  1.688 &  0.725 &  0.341 & -0.531 & -0.529 & -0.502 & -0.338 \\
4  & AR-first & Video & -0.369 & -0.542 & -0.433 & -0.520 & -0.368 & -0.361 & -0.529 & -0.513 \\
8  & AR-first & AR    &  1.359 &  0.103 &  0.606 &  0.439 & -0.407 & -0.496 & -0.017 & -0.271 \\
8  & AR-first & Video & -0.309 & -0.098 & -0.273 & -0.408 & -0.210 &  0.391 & -0.248 & -0.314 \\
10 & AR-first & AR    &  0.800 &  0.432 &  0.136 &  0.072 &  0.056 &  0.221 & -0.259 & -0.175 \\
10 & AR-first & Video & -0.242 & -0.311 & -0.219 & -0.372 &  0.080 & -0.148 &  0.072 & -0.163 \\
12 & AR-first & AR    & -0.057 &  0.653 &  0.220 & -0.169 & -0.487 & -0.520 & -0.549 & -0.408 \\
12 & AR-first & Video & -0.362 & -0.370 & -0.444 & -0.500 & -0.539 & -0.440 & -0.462 & -0.483 \\
14 & AR-first & AR    &  1.014 &  0.735 & -0.264 &  0.061 & -0.501 & -0.509 & -0.556 & -0.469 \\
14 & AR-first & Video & -0.441 & -0.510 & -0.466 & -0.488 & -0.513 & -0.554 & -0.315 & -0.474 \\
16 & AR-first & AR    & -0.241 &  0.163 & -0.344 & -0.497 & -0.562 & -0.545 & -0.564 & -0.510 \\
16 & AR-first & Video & -0.477 & -0.390 & -0.517 & -0.485 & -0.469 & -0.497 & -0.456 & -0.443 \\
18 & AR-first & AR    &  0.526 &  0.676 &  1.256 &  0.742 &  0.542 & -0.346 &  0.651 &  0.138 \\
18 & AR-first & Video & -0.013 &  0.295 &  0.094 &  0.074 &  0.246 &  0.295 &  0.154 &  0.292 \\
20 & AR-first & AR    &  1.822 &  1.841 &  0.565 &  0.635 & -0.458 & -0.483 & -0.464 & -0.011 \\
20 & AR-first & Video &  0.071 &  0.553 &  0.110 & -0.086 & -0.228 & -0.500 & -0.359 & -0.217 \\
22 & AR-first & AR    &  1.218 &  1.556 &  0.444 &  0.478 & -0.264 & -0.495 & -0.491 & -0.510 \\
22 & AR-first & Video & -0.189 & -0.150 &  0.026 &  0.096 & -0.156 &  0.085 & -0.049 & -0.154 \\
24 & AR-first & AR    &  0.613 &  0.278 & -0.060 & -0.136 & -0.468 & -0.505 & -0.523 & -0.350 \\
24 & AR-first & Video & -0.267 & -0.243 & -0.383 & -0.268 & -0.393 & -0.402 & -0.388 & -0.301 \\
25 & AR-first & AR    &  0.919 &  0.712 &  0.178 & -0.086 & -0.447 & -0.411 & -0.431 & -0.137 \\
25 & AR-first & Video & -0.211 & -0.252 &  0.079 & -0.006 & -0.175 & -0.357 & -0.255 & -0.331 \\
\bottomrule
\end{tabular}}
\label{tab:participant_zscores_combined}
\end{table*}


\end{document}